\documentclass[nolinenumbers]{aastex631}

\begin{document}

\title{Two Possible Orbital Histories of Phobos}

\correspondingauthor{Matija \'Cuk}
\email{mcuk@seti.org}

\author[0000-0003-1226-7960]{Matija \'Cuk}
\affiliation{SETI Institute, 339 N Bernardo Ave, Suite 200, Mountain View, CA 94043, USA} 

\author[0009-0007-2467-0139]{Kaustub P. Anand}
\affiliation{Department of Physics and Astronomy, \\
Purdue University, \\
525 Northwestern Ave, West Lafayette, IN 47907, USA}

\author[0000-0003-1656-9704]{David A. Minton}
\affiliation{Department of Physics and Astronomy, \\
Purdue University, \\
525 Northwestern Ave, West Lafayette, IN 47907, USA}
\affiliation{Department of Earth, Atmospheric and Planetary Sciences \\
Purdue University  \\
550 Stadium Mall Drive\\ 
West Lafayette, IN 47907, USA}

\begin{abstract}
The two moons of Mars, Phobos and Deimos, have orbits that are close to martian equator, indicating their formation from a circumplanetary disk. Phobos is currently migrating toward Mars due to tidal dissipation within the planet, and may be disrupted into a ring in few tens of Myr. The past evolution of Phobos is not fully understood, with one possibility being that Phobos formed in the early Solar System just interior to the synchronous orbit and migrated inward over several Gyr. Alternatively, Phobos may be the most recent product of an ongoing martian ring-moon cycle that lasted several Gyr but formed Phobos only 100 Myr ago at the fluid Roche limit. Here we use numerical integrations to simulate past evolution of Phobos in both of these scenarios and test whether Phobos's small eccentricity and inclination are consistent with either of these hypotheses. During its tidal evolution, Phobos crossed multiple resonances with both the rotation of Mars and the apparent motion of the Sun, requiring detailed numerical modeling of these dynamical events. Furthermore, Phobos crossed resonances with Deimos, which can affect the orbit of Deimos. We find that both the ancient Phobos hypothesis and the ring-moon cycle are fully consistent with the present orbit of Phobos, with no currently available means of distinguishing between these very different dynamical histories. Furthermore, we find that Deimos is affected by weak chaos caused by secular resonances with the planetary system, making the eccentricity of Deimos an ineffective constraint on the past migration of Phobos. 
\end{abstract}


\section{Introduction} \label{sec:intro}

The origin of two martian satellites Phobos and Deimos has long been a mystery. While the idea that Phobos and Deimos are captured asteroids has been historically widespread, their orbits close to Mars's equatorial plane (Table \ref{moons}) suggest their origin from a circum-martian disk \citep{bur92, hyo22}. A giant impact on Mars has been a leading candidate for the origin of this disk \citep{cra11, ros16, cit15, can18}, it was recently proposed that this circumplanetary disk consisted of captured asteroidal material \citep{keg24}. Alternatively, \citet{bag21} proposed that Phobos and Deimos were formed on initially highly-eccentric but near-equatorial orbits as collisional fragments of a single moon, but this hypothesis requires implausible ejection velocity fields and highly divergent tidal properties of Phobos and Deimos \citep{hyo22}.

\begin{deluxetable}{cccccc}
\tablenum{1}
%\vspace{-.2in}
\tablecaption{Physical and orbital parameters of Phobos and Deimos. Radii and masses are from \citet{md99}, and the mean orbital elements are from \citet{jac14} (Mars radius was taken to be $R_M=3394$~km). \label{moons}}
%\tablewidth{0pt}
\tablehead{\colhead{Moon} & \colhead{Radius [km]} & \colhead{Mass [$10^{15}$~kg]} & \colhead{Semimajor axis [$R_M$]} & \colhead{Eccentricity} & \colhead{Inclination [$^{\circ}$]}}
%\decimalcolnumbers
\startdata 
Phobos & 11 & 10.8  &  2.762 & 0.01511 & 1.076\\
Deimos & \phantom{0}6 & \phantom{0}1.8 & 6.912 & 0.00027 & 1.789\\
\enddata
%\tablecomments{}
\end{deluxetable}

\begin{figure}[ht]
\epsscale{.6}
\plotone{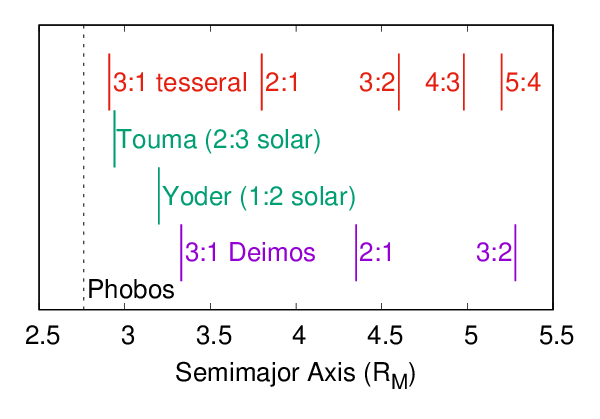}
\caption{Different resonances potentially encountered by Phobos in the past. Tesseral resonances with the rotation of Mars are plotted in red, semi-secular resonances with the Sun in teal, and mean-motion resonances with Deimos in purple. While the tesseral and Deimos resonances are labeled by the ratio of Phobos's mean motion to the martian rotation rate and Deimos's mean motion, respectively, semi-secular resonances are quantified by the ratio of Phobos's apsidal precession rate to the martian mean motion. The dashed line shows the present semimajor axis of Phobos.}
\label{fig1}
\end{figure}

If we accept that Phobos and Deimos formed on circular, equatorial orbits around Mars, we need to address the question of their orbital evolution since formation. Phobos is currently spiraling in toward Mars \citep{jac14}, and is expected to be tidally disrupted within tens of Myrs \citep{bla15}. As Phobos is interior to the synchronous orbit, it is assumed that it formed further out than it is now, but within the synchronous radius at about 6 Mars radii ($R_M$), and then migrated to its present semimajor axis at 2.76~$R_M$ \citep[see Table \ref{moons};][]{bur92}. On the other hand, smaller and more distant Deimos was unlikely to have migrated much due to tides raised on Mars over the age of the Solar System. \citet{yod82} examined the dynamical history of Phobos in some detail and found that Phobos likely crossed resonances with martian rotation (i.e. tesseral resonances), as well as resonances with the Sun and Deimos (Table \ref{resonances}). While the tesseral and solar resonances could plausibly explain the present eccentricity and inclination of Phobos, \citet{yod82} found that Deimos should have a much higher eccentricity than observed if Phobos formed beyond Deimos's interior 2:1 mean-motion resonance (MMR) at 4.35~$R_M$ (Fig. \ref{fig1}).

\begin{deluxetable*}{lcccc}
\tablenum{2}
\vspace{-.2in}
\tablecaption{Resonances potentially encountered by Phobos in the past that were considered in this paper. The resonant argument associated with the largest resonant Hamiltonian term is labeled as ``primary'', and the second most important ``secondary''. Only arguments relevant to \citet{yod82} or our work are given, and lack of secondary arguments in the table does not imply they do not exist. $\lambda$ is the mean longitude, $\varpi$ and $\Omega$ are the longitudes of pericenter and ascending node, while $\Omega_{Eq}$ and $\phi_M$ are Mars's equinox and rotation angle. Subscripts P, D, and M refer to Phobos, Deimos and Mars, respectively.\label{resonances}}
\tablewidth{0pt}
\tablehead{ \colhead{Resonance} & \colhead{$a_P$ ($R_M$)} & \colhead{Primary argument} & \colhead{Secondary argument} & \colhead{Paper Section}}
%\decimalcolnumbers
\startdata 
3:1 tesseral & 2.91  &  $3 \phi_M - \lambda_P - 2 \Omega_P$ & $3 \phi_M - \lambda_P - 2 \varpi_P$ & \ref{sec:31res}\\
Touma (2:3 solar) & 2.94 &  $2 \varpi_P - 2 \lambda_M - \Omega_P + \Omega_{Eq}$ & & \ref{sec:touma}\\
Yoder (1:2 solar) & 3.2\phantom{0} &  $2 \varpi_P - \lambda_M - \varpi_M$ & $2 \varpi_P - 2 \lambda_M - 2\Omega_P + 2\Omega_{Eq}$ & \ref{sec:yoder}\\
3:1 with Deimos & 3.33 & $3 \lambda_D-\lambda_P - \varpi_P - \varpi_D$ &  & \ref{sec:deimos}\\
2:1 tesseral & 3.8\phantom{0} & $2 \phi_M - \lambda_P - \varpi_P$  &  $4 \phi_M - 2 \lambda_P - 2 \Omega_P$ & \ref{sec:2to1} \\
2:1 with Deimos & 4.35 & $2 \lambda_D-\lambda_P - \varpi_D$ &  & \ref{sec:deimos} \\
3:2 tesseral & 4.6\phantom{0} & $3 \phi_M - 2 \lambda_P - \varpi_P$  &  $6 \phi_M - 4 \lambda_P - 2 \Omega_P$ & \ref{sec:early} \\
4:3 tesseral & 4.98 & $4 \phi_M - 3\lambda_P - \varpi_P$  &  $8 \phi_M - 6 \lambda_P - 2 \Omega_P$ & \ref{sec:early} \\
5:4 tesseral & 5.2\phantom{0}& $5 \phi_M - 4 \lambda_P - \varpi_P$ & & \ref{sec:early}\\
\enddata
%\tablecomments{}
\end{deluxetable*}

A fundamentally different past history of Phobos was proposed by \citet{hes17}. They note that formation of martian moons from a disk that is largely interior to synchronous orbit implies the formation of multiple generations of moons, each more massive than Phobos. Due to inward tidal evolution interior to synchronous orbit, these moons would each have suffered tidal disruptions into a ring, meeting the same fate expected for Phobos. Observations of Saturn's ring-moon system imply that satellites form at the outer edge of the rings and then these satellites tend to evolve outward due to gravitational interaction with the ring itself \citep{cha10}. Therefore \citet{hes17} recognized that ring formation is not an end of the system's evolution, but that the ring should give rise to a new, smaller, moon that will initially migrate outward through interacting with the ring. Eventually, as the ring becomes depleted by losing material to the planet, the new moon will switch from outward to inward migration due to tides raised on Mars, as would still be well inside the synchronous radius. \citet{hes17} proposed that this cycle of formation of moons from rings and destruction of moons into rings may have repeated itself multiple times over the history of the system, with every new moon being less massive that the previous one. The number of cycles and the reduction in the moon's mass from one cycle to the next are determined by physical properties of the hypothetical rings and moons, but reasonable assumptions lead to roughly 5 to 7 cycles over the age of Mars, with each new moon being smaller by a factor of five in mass relative to the previous one. In the view of \citet{hes17}, Phobos is just the latest incarnation of this ring-moon cycle, and may have formed within the last 100-200~Myr at fluid Roche limit (FRL) at about 3.2~$R_M$, migrated outward while a massive ring was present, and has since been migrating inward. 

The hypothesis of ring-moon cycles at Mars may help explain the previously mysterious high inclination and low eccentricity of Deimos. Most dynamical mechanisms, such as planetesimal flybys \citep{nes14, pah15, bra20} tend to excite both eccentricity and inclination. While eccentricity is more readily damped by tides within the satellite than inclination, the large semimajor axis of Deimos makes damping of its eccentricity implausible. Also, if we assume that Deimos has physical properties comparable with Phobos, which is on a much closer yet notably eccentric orbit, large scale damping of Deimos's eccentricity appears unlikely. \citet{cuk20} show that a convergent 3:1 MMR between a past inner moon of Mars with mass $m=18\ m_{\rm Phobos}$ and Deimos would leave Deimos with correct high inclination and low eccentricity. The asymmetry between the inclination and eccentricity effects of the resonance is caused by the presence of the Sun, which induces a forced inclination of Deimos of about one degree, making resonance capture much more likely \citep{cuk20}. The 3:1 MMR with Deimos is located at 3.33~$R_M$ and is expected to be crossed both during outward and inward migration of past massive inner moons of Mars \citep{hes17}. 

Other data has been put forward in support of recent formation of Phobos from a ring. \citet{hu20} find that the overall shape of Phobos is best matched by tidal forces it would have experienced at 3.3~$R_M$, indicating a formation close to that distance, which is very close to the FRL. On the other hand, while \citet{mad23} find that the ring-moon cycles at Mars are plausible, they also argue that recent formation of Phobos from a ring is not consistent with no trace of this ring surviving to the present. \citet{mad23} therefore propose that the ring-moon cycle at Mars, if it operated, terminated more than a Gyr ago, and that Phobos is not its latest incarnation, but an initially distant moon that has been migrating inward over the age of the system.

The question of timing and location of the formation of the martian moons is further complicated by the possibility of a {\em sesquinary catastrophe}. This mechanism was proposed by \citet{cuk23} and involves a runaway erosion of a close-in planetary moon by sesquinary impacts, which are caused by ejecta that escape from a satellite but remain in a planetocentric orbit to eventual re-impact the same satellite \citep{zah08}. For sesquinary impacts to become erosive, the eccentricity and/or inclination of the moon must be large enough so that ejecta on orbits that are initially similar to that of the moon but subsequently precess out of alignment can re-impact the moon at velocities much higher than the satellite's own escape velocity. For Phobos and Deimos, based on estimates of \citet{cuk23}, it is likely that $e \simeq 0.1$ or $\sin{i} \simeq 0.1$ should be enough to cause a sesquinary catastrophe and cause the disruption of the moon. It is expected that the moon would re-accrete on a circular and planar orbit that would have the same angular momentum as the original moon. 

Deimos and Phobos, if it is ancient, are expected to have interacted strongly with the short-lived massive close-in satellites of Mars that were also produced in the event that formed the original protosatellite disk, such as the giant impact \citep{ros16}. Strong resonances with the large inner moons may have greatly excited the orbits of the outermost satellites, but dynamical excitation beyond the threshold of sesquinary catastrophe would lead to re-accretion on a circular and planar orbit. In case of the ring-moon cycle proposed by \citet{hes17}, these re-accretion events could have happened well after the epoch of the first generation of moons, as long as the martian rings were able to produce new moons large enough to excite the orbits of more distant satellites. The survival of Deimos' inclination likely originated in one such resonance. \citet{cuk20} indicates that Deimos was not re-accreted over the last 3~Gyr or more. As Phobos may have experienced much more recent orbital resonances with the martian rotation and the Sun, it is harder to put any constraints on its age. 

In this paper we will model the past orbital evolution of Phobos, much like \citet{yod82} did, but using direct numerical simulations not available to them. We will also explore both hypotheses for the origin of Phobos: that Phobos (re)formed several Gyr ago relatively close to the synchronous orbit (the ``ancient Phobos" hypothesis, APH) or that Phobos formed from a martian ring around 3.2~$R_M$ on the order of 100~Myr ago (the ``recent Phobos" hypothesis, RPH). We will also discuss a range of possible past tidal behaviors of Mars, including the question of whether the tidal response of Mars follows the ``constant $Q$'' or ``constant lag'' frequency dependence, or the tidal $Q$ may even increase with frequency \citep{bag21}. This issue of frequency dependence is less important for resonance crossings than for determining how close to the synchronous orbit Phobos may have formed, assuming APH. We assume a tidal torque on Phobos in the plane of the orbit corresponding to martian $Q=80$ and $k_2=0.14$, and these will be our nominal tidal parameters throughout the paper unless specified otherwise.

To be successful, both hypotheses for the formation of Phobos must result in the present-day orbit of Phobos, therefore we will address important dynamical events, essentially resonances or different types, in the reverse chronological order. While our integrations will have to be forward in time in order to accurately model resonance crossings, we will study the more recent resonances first, so that the initial conditions we prefer for one resonance can be the desired outcome for the next resonance to be modeled.     

\section{Methods}\label{sec:methods}

In our numerical experiments we exclusively used different variants of the mixed variable symplectic integrator {\sc simpl} \citep[Symplectic Integrator for Moons and PLanets]{cuk16} that is directly based on the algorithm of \citet{cha02}. The main feature of {\sc simpl} is that it integrates both the planetary system and the satellite system of one of the planets, and it can fully account for the perturbations of the Sun and the planets (other than the parent planet) on the satellites' orbits. While both direct and indirect perturbations of the planets can be modeled in {\sc simpl}, unless otherwise stated, we have ignored direct perturbations of the planets on the moons to save computational time, and included only the perturbations of the other planets on the orbit (and sometimes the spin) of Mars. This approach is justified outside very rare direct resonances between a satellite of one planet and another planet in the system \citep{kau76, yod76, cuk07}. Additionally, {\sc simpl} includes simple treatments of tidal torques on the satellites and can also include artificial migration of the satellites through an artificial torque \citep{cuk16}, which we use to simulate past ring torques on Phobos. 

\begin{deluxetable}{lll}
\tablenum{3}
%\vspace{-.2in}
\tablecaption{Summary of numerical integrators used in this paper. All the derivatives of {\sc simpl} constain all of its basic capabilities. \label{integrators}}
%\tablewidth{0pt}
\tablehead{\colhead{Integrator} & \colhead{Description} & \colhead{Reference}}
%\decimalcolnumbers
\startdata 
{\sc simpl} & Includes tides, $J_2$, moon-moon, solar and planetary perturbations & \citet{cuk16}\\
{\sc simpl-harm} & Includes perturbations from tesseral gravity harmonics & This work\\
{\sc s-simpl} & Includes precessing spin axis of Mars due to planetary torques & \citet{cuk18}\\
{\sc s-simpl-harm} & Combines {\sc s-simpl} and {\sc simpl-harm} & This work\\
{\sc s-simpl-read} & {\sc s-simpl} with pre-integrated planetary orbits and martian spin precession & This work\\
\enddata
%\tablecomments{}
\end{deluxetable}

Simulations presented in this paper required some modifications to {\sc simpl}. The most important modification was introduction of perturbations from gravity harmonics of Mars. These were introduced as acceleration ``kicks'' in Cartesian coordinates. The values for the strength of martian gravitational harmonics is taken from \citet{kon20}. Given significant computational cost of the inclusion of the harmonics, we often included only the relevant gravitational harmonics in simulations of resonance crossing. Overall, in addition to $J_2$ which is hardwired in {\sc simpl}, our model includes harmonics $C_{22}$, $C_{32}$, $C_{33}$, $C_{42}$, $C_{43}$, $C_{44}$, $C_{54}$ and $C_{55}$ (as well as corresponding sine harmonics $S_{22}$ etc.). We refer to the version of {\sc simpl} that includes gravitational harmonics as {\sc simpl-harm} (Table \ref{integrators}). Another modification of  {\sc simpl} we used is the direct integration of the variations of martian obliquity due to planetary perturbations \citep{war73, tou93, las04}. The changing obliquity of Mars is relevant for some of Phobos's semi-secular resonances with the Sun, such as the Touma resonance (Table \ref{resonances}). This version of {\sc simpl} has already been used by \citep{cuk18} to integrate the spin dynamics of Saturn due to its spin-orbit resonance with the planetary system, and we will refer to it as {\sc s-simpl} (S for spin). Two additional variants of the integrator used here are {\sc s-simpl-harm} which simply combines these two additional features, including both the spin dynamics of Mars and perturbations from its gravitational harmonics on the moons. Finally, for some very long-term integrations involving Deimos and the planets we introduce the variant {\sc s-simpl-read} in which the planets and the satellites are integrated separately, which we describe detail in Section \ref{sec:deimos}.

\section{3:1 Tesseral Resonance}\label{sec:31res}

\begin{figure}[ht]
\epsscale{.7}
\plotone{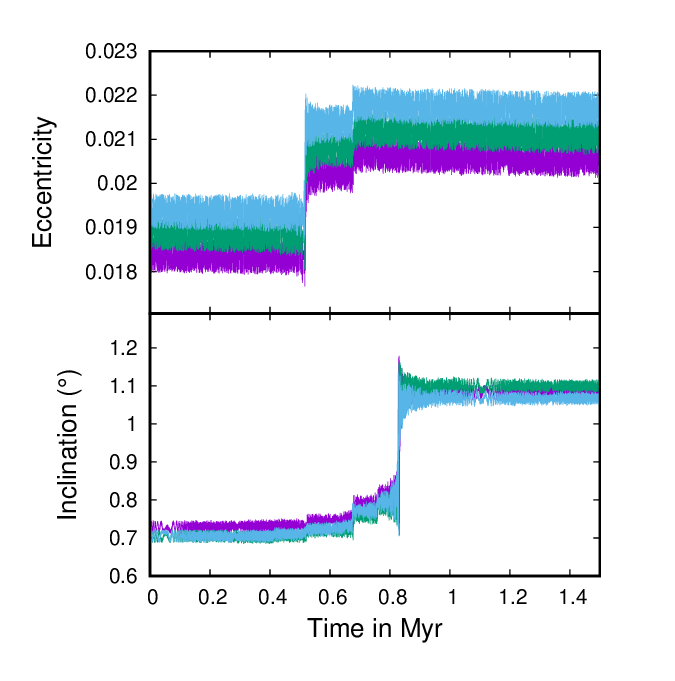}
\caption{Three different numerical simulations of Phobos's passage through the 3:1 tesseral resonance with the rotation of Mars at 2.9~$R_M$. The slightly different initial conditions were chosen so to result in the present orbit of Phobos after this resonance. The changes in Phobos's $e$ and $i$ during this resonance passage are consistent across many simulations, enabling us to estimate Phobos's pre-crossing orbital parameters with confidence. Under our standard model assumptions ($Q_M=80, k_{2M}=0.14$), this crossing happened about 16~Myr ago.}
\label{31harm}
\end{figure}

As \citet{yod82} found, the 3:1 resonance between the mean motion of Phobos and the rotation of Mars is the last orbital resonance crossed by Phobos (Table \ref{resonances}. It is primarily driven by the $C_{33}$ and $S_{33}$ tesseral harmonics of Mars. In our model we also include the $C_{43}$ and $S_{43}$ terms in the martian gravity field, but we ignore all others. We also assume Mars is on a fixed orbit with an eccentricity of $e_M=0.09$ and a fixed axial tilt of $\epsilon=25^{\circ}$, and we have found no indication that the Sun would affect the dynamics of this resonance. 

We conducted a number of integrations with a range of initial conditions, and we found that the 3:1 tesseral resonance almost always leads to a jump in the inclination and eccentricity of Phobos. The main kicks in inclination and eccentricity closely match the estimates of $\Delta i=0.28^{\circ}$ and $\Delta e=0.002$ from \citet{yod82}. The actual change in  inclination and eccentricity was slightly larger due to minor subresonances of the 3:1 tesseral resonance. In particular the noticeable jump in eccentricity and the corresponding jump in inclination at about $0.7$~Myr in Fig. \ref{31harm} are caused by the $C_{43}$ and $S_{43}$ harmonics (and disappear when these terms are removed from the model). 

As demonstrated by Fig. \ref{31harm}, all of the kicks in $i$ and $e$ are relatively stable from simulation to simulation, with only a minor stochastic component. The only uncertainty regarding this resonance crossing comes from the $<$10\% of cases in which there is a short-lived capture into the strongest sub-resonance of the 3:1 tesseral resonance. Permanent capture into tesseral resonances requires migration toward the synchronous orbit, which is opposite to the direction of tidal evolution and can therefore be excluded for Phobos \citep{yod82}. However, chaotic interactions allow for temporary residence in resonance, and as Phobos is migrating inward it must move to lower inclinations to stay in the resonance. During this phase, inclination decreases, reducing the size of the inclination kick and sometimes even ending up being lower than before the resonant encounter. Given the low probability nature of this outcome (especially in contrast to large-scale uncertainties for other resonances) we will ignore it here, especially as there is no direct implication for distinguishing between APH and RPH.

Given the relatively consistent range of outcomes from the 3:1 tesseral resonance, we concentrated our later simulations on reproducing the current orbit of Phobos, as illustrated in Fig.~\ref{31harm}. When reconstructing a nominal orbital history for Phobos, we need to take into account both the effects of resonance crossings and the evolution of eccentricity between the resonances. Both the tides raised on Mars by a super-synchronously orbiting Phobos and tides raised by Mars on a synchronously rotating Phobos act to decrease Phobos's eccentricity. In case of planetary tides this is a side-effect of orbital contraction, and, assuming constant-$Q$ tides for Mars, the two are correlated as $e \propto a^{3.1}$, as found by \citet{yod82} and confirmed in our numerical simulations. 

The contribution of satellite tides to the damping of Phobos's eccentricity is less clear and depends on the material properties of Phobos, with the plausible effect ranging from negligible to being comparable to that of martian tides \citep{yod82}. For purposes of setting up simulations, we will assume that eccentricity damping by satellite tides can be estimated using the approach of \citet{gol09}, who predict $Q_P=100$ and $k_{2P}=1.1 \times 10^{-4}$ for $R=11$~km rubble-pile Phobos. Using the relation from \citet{can99} Eq. 4 modified for Phobos, we have:
\begin{equation}
\Bigl( {de \over dt}\Bigr) \simeq {e \over a} \Bigl( { da \over dt} \Bigr) \Bigl( 3.1 + {7 \over 2} A \Bigr)
\end{equation}
where the first term in parentheses is due to martian tides and the second is due to Phobos tides, and we assumed that all of semimajor axis evolution is due to tides within Mars. The parameter $A$ is defined as:
\begin{equation}
A = {k_{2P} Q_M \over k_{2M} Q_P} \Bigl( {\rho_M \over \rho_P} \Bigr)^2 {R_M \over R_P}
\end{equation}
where $\rho$ is mean density and subscripts M and P refer to Mars and Phobos. If we use $k_{2M}=0.15$ and $Q_M=95$\footnote{The discrepancy from values $k_{2M}=0.14$ and $Q_M=80$ used in our simulations is minor, and here we prefer using more modern values.} we get $A=0.96$, implying that the eccentricity varies as $\propto a^{3.36}$ between resonances due to satellite tides alone, the combined damping due to both tidal effects results in $e \propto a^{6.46}$. Using this relation, the eccentricity of Phobos should have been about $e_P=0.021$ after the 3:1 tesseral resonance. Numerical simulations such as those shown in Fig.~\ref{31harm} indicate that the pre-resonance orbit of Phobos most likely had $e_P \approx 0.019$ and $i_P \approx 0.7^{\circ}$. This enables us to move further back in time and consider the last solar resonance that Phobos has crossed during its tidal evolution.

\section{Touma (2:3) Semi-Secular Solar Resonance}\label{sec:touma}

Here we address the penultimate resonance encountered by Phobos that involves a semi-secular interactions between Phobos and the Sun. ``Semi-secular'' means that the precession of Phobos is in resonance with the apparent mean motion of the Sun (i.e. the heliocentric mean motion of Mars). The most important semi-secular resonance is the evection resonance \citep{tou98}, in which the moons precession period is equal to the planet's orbital period, and the argument $2\lambda_{planet}-2\varpi_{moon}$ is librating ($\lambda$ being the mean longitude and $\varpi$ the longitude of pericenter). However, Phobos yet has to encounter the evection resonance proper \citep{yok05}. 

\citet{yod82} correctly identified that Phobos should have crossed the 2:1 harmonic of evection when Phobos was at $a=3.2~R_M$ and its precession period was two martian years; this ``Yoder'' resonance will be addressed in Section \ref{sec:yoder}. Subsequently, \citet{tou98}, while studying the early evolution of the Earth-Moon system, identified another harmonic of evection that happens when the orbital precession rate of a moon is about two thirds of the planet's mean motion. This is a mixed eccentricity-inclination semi-secular resonance with the argument $2\lambda_{planet}-2\varpi_{moon}+\Omega_{moon}-\Omega_{eq}$, where $\Omega$ is the longitude of the ascending node and $\Omega_{eq}$ is the longitude of the planet's vernal equinox. \citet{tou98} called this resonance ``eviction'', that is ``evection with an $i$''. In order to avoid confusion with other semi-secular resonances involving inclination \citep[for examples see][]{yok05}, we will call this resonance the Touma resonance.  

\begin{figure}[ht]
\epsscale{.6}
\plotone{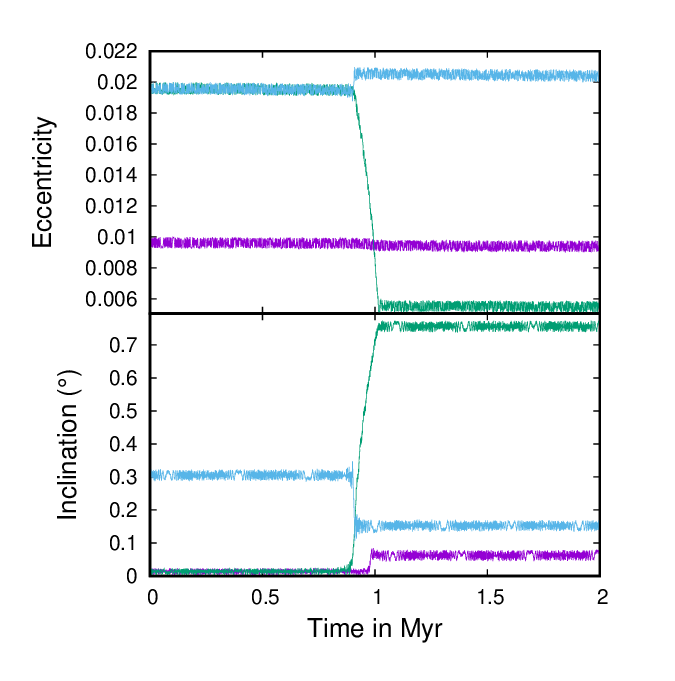}
\caption{Three different numerical simulations of Phobos's passage through the Touma (2:3) semi-secular resonance with the Sun at 2.94~$R_M$, in a model with a fixed orbit of Mars. The purple line shows a case with initial $i=0$ that has eccentricity too low for resonance capture. The green line plots the case with higher initial $e$ that results in resonance capture. A case with the same initial $e$ but also substantial $i$ also avoids capture (light blue). Using standard assumptions, this crossing happened about 20~Myr ago.}
\label{32evi}
\end{figure}

A known characteristic of the Touma resonance is that it can capture the orbit of a satellite that is migrating inward, which makes it applicable to Phobos. As this is a secular resonance from the moon's point of view, Phobos keeps migrating inward after the resonance capture following the trajectory of constant precession rate in the orbital element space. This means that the inclination increases as semimajor axis decreases. As is usual with resonance capture, trapping into the Touma resonance requires low initial inclination of Phobos \citep[cf][]{tou98}. The Touma resonance is a mixed eccentricity-inclination resonance, and its Hamiltonian is proportional to $e^2_P i_P$. Therefore the probability of capture increases with larger initial eccentricity. Depending on the initial orbital elements of Phobos as it encounters the Touma resonance, Phobos can either experience capture into the resonance or a ``jump'' through the resonance.    

Figure~\ref{32evi} plots three simulations that illustrate two distinct modes of crossing the Touma resonance at 2.94~$R_M$. In these simulations Mars was on a fixed Keplerian orbit with $e_M=0.09$ and had a constant axial tilt of $\epsilon=25^{\circ}$. The purple track shows the orbital elements of Phobos crossing the Touma resonance while migrating inward (assuming martian $Q=80$, $k_2=0.14$), on an initially planar orbit with $e_P=0.01$. In this case there is no resonant capture, but these is only a small kick to both eccentricity and inclination. The green curve plots a simulation that is identical except that we initially set $e_P=0.02$. The outcome is very different, as there is capture into the Touma resonance, with eccentricity dropping to $e_P< 0.006$ and inclination rising to $i_P > 0.7^{\circ}$. Capture becomes difficult for higher initial inclinations, as shown by the cyan curve, which also starts with $e_P=0.02$ but now has initial $i_P=0.3^{\circ}$, and results in a resonant jump rather than resonance capture.  

The two distinct modes of passing through the Touma resonance, jump and capture, have crucial implications for reconstructing the past orbital evolution of Phobos. In the context of APH,  based on results of \citet{yod82} we can expect that Phobos had a significant inclination after crossing the 2:1 tesseral resonance at $3.8~R_M$. In that case we could expect Phobos to cross the Touma resonance with a jump and without a capture. On the other hand, according to RPH, Phobos never ventured beyond $3.3~R_M$ so it could in principle encounter Touma resonance with a low inclination, possibly leading to capture. Note that the last assumption depends on the details of Phobos's interaction with the Yoder (or 1:2) solar resonance at $3.2~R_M$, which is addressed in the next section. In any case, to correctly reconstruct the orbit of Phobos preceding the Touma resonance we need to consider the two distinct pre-resonance paths and follow both in detail.

One simplifying factor when dealing with the Touma resonance is that it affects eccentricity and inclination in closely related ways. As the resonance argument is $2\lambda_{planet}-2\varpi_{moon}+\Omega_{moon}-\Omega_{eq}$, Laplace-Lagrange equations \citep{md99} suggest that $e {\dot e}= - 2 \sin{i}(di/dt)$. This in turn means that the quantity $C_T= e^2 + 2 \sin^2{i}$ is conserved for small $e$ and $i$. We can use this property of the resonance to set up initial conditions in $e$ and $i$ that can evolve into our nominal post-resonance target of $e_P \approx 0.02$ and $i_P \approx 0.7^{\circ}$. 

\begin{figure}[ht]
\epsscale{.6}
\plotone{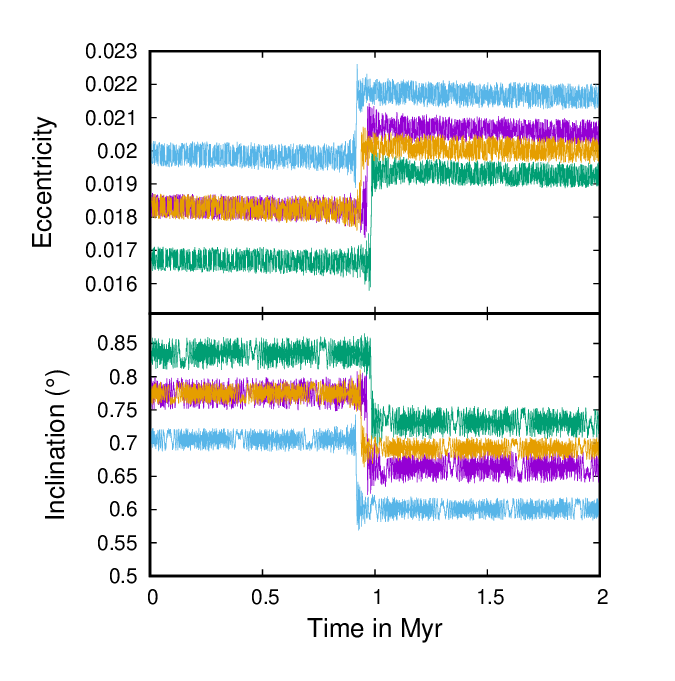}
\caption{Four different numerical simulations of Phobos's non-capture crossing through the Touma (2:3) semi-secular resonance with the Sun at 2.94~$R_M$, with the post-resonance orbit of Phobos consistent with our nominal orbital history. We can conclude that one set possible pre-resonance orbital elements of Phobos is approximately $e_P = 0.018$ and $i_P = 0.8^{\circ}$.}
\label{32back}
\end{figure}

Figure \ref{32back} shows four different simulations of Touma resonance crossing that experienced jump rather than capture. In the two of the simulations (magenta and green) the axial tilt of Mars was fixed at $\epsilon=45^{\circ}$, while the cyan and gold lines plot simulations with $\epsilon=25^{\circ}$; Mars was assumed to have a constant obliquity and eccentricity in these simulations. While the jumps in $e_P$ and $i_P$ are noticeable, the pre-resonant orbit of Phobos only had to be slightly less eccentric and more inclined than the post-resonance elements to get a rough match. These pre-resonance elements will serve as a guide to our further exploration of Phobos's dynamical history.

\begin{figure}[ht]
\epsscale{1.15}
\plottwo{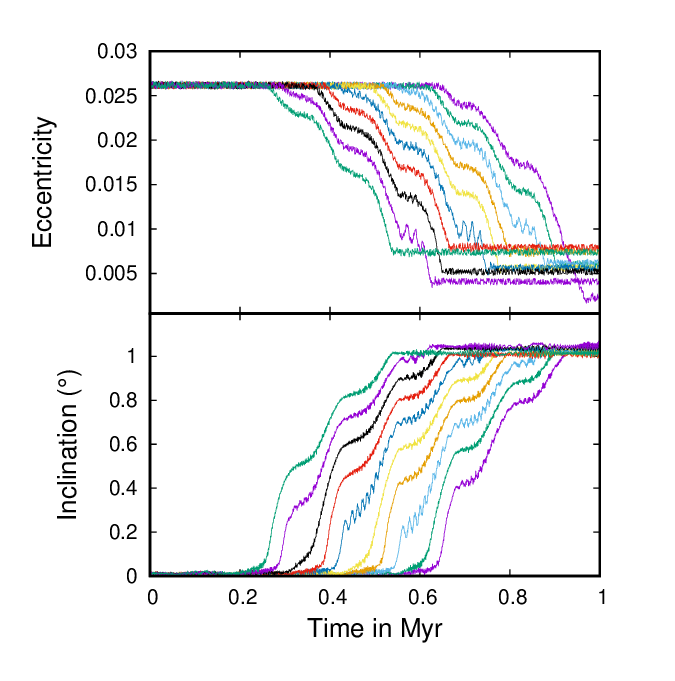}{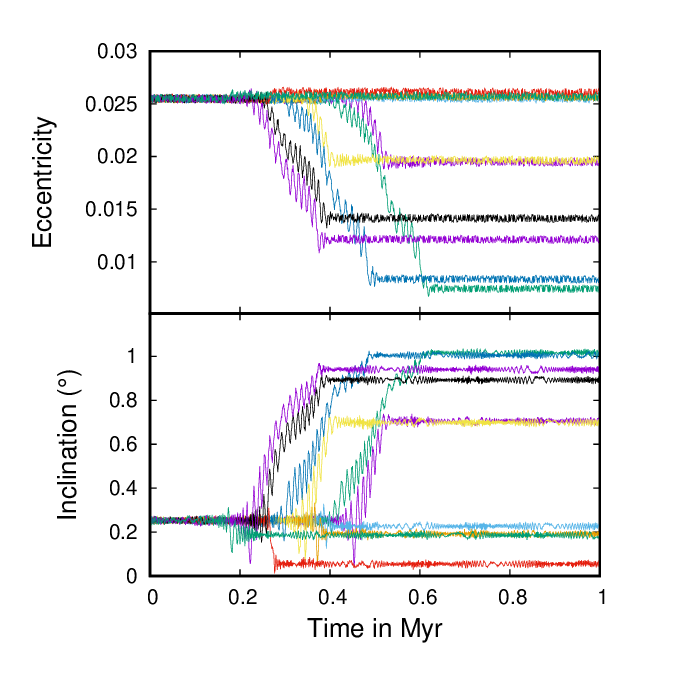}
\caption{Two sets of ten different numerical simulations of Phobos's capture the Touma (2:3) semi-secular resonance with the Sun at 2.94~$R_M$, with conserved quantity $C_T=e_P^2+2\sin^2{i_P}$ consistent with our nominal orbital history. The simulations on the right have initial $i_P=0$ and martian tidal $Q=120$ while those on the left have initial $i_P=0.25^{\circ}$ and martian $Q=80$. In each case, ten different simulations each had slightly different initial semimajor axis for Phobos, resulting in different times of resonance capture. While the initially planar orbits almost all evolve to eccentricities below $e_P=0.01$, simulations with a non-zero initial inclination drop out of resonance earlier. Our nominal reconstructed outcome of $e_P \approx 0.02$ and $i_P \approx 0.7^{\circ}$ being within the spread of outcomes.}
\label{32ssimpl}
\end{figure}

Capture into the Touma resonance offers an alternate route to the present orbit of Phobos that starts with a lower pre-resonant inclination. A completely planar pre-resonance orbit of Phobos can be ruled out, as Touma resonance capture typically does not break until $e_P< 0.01$, which is inconsistent with Phobos's subsequent tidal evolution (green line in Fig. \ref{32evi}). We explored a number of dynamical effects that could lead to an earlier breaking of the Touma resonance. Simply using a smaller obliquity of Mars did not have significant effect. 

Next we moved to using {\sc s-simpl}, which directly integrates the obliquity variations of Mars. The left-hand panels in Fig. \ref{32ssimpl} shows ten simulations (made using {\sc s-simpl}) of capture into the Touma resonance with initially equatorial orbits of Phobos ($i_P=0^{\circ}$), with each simulation starting with slightly different distance from Mars. The initial martian axial tilt was $\epsilon=25^{\circ}$, and in these simulations we used $Q=120$ for Mars in order to make the resonance capture longer and have Mars undergo more obliquity cycles during that time. Despite an obvious effect of the martian obliquity on the Touma resonance (evident as ``wavines'' of the $e$ and $i$ curves), and a larger scatter of outcomes, the resulting orbits are not consistent with the subsequent evolution of Phobos down to the present. 

On the right-hand side in Fig. \ref{32ssimpl}, we show an identical set of simulations but with a pre-resonant inclination of Phobos of about $i_P=0.25^{\circ}$. We find that a pre-resonant $i_P=0.25^{\circ}$ makes the Touma resonance capture less stable, resulting in resonance breaking at higher $e_P$ and lower $i_P$ than in previous examples. The outcomes from this set of simulations include the range of post-resonance elements of Phobos we were aiming for.    

We conclude that the orbit of Phobos before Touma resonance could have possessed two separate and distinct combinations of eccentricity and inclination. Non-capture passage through the Touma resonance implies pre-resonant $e_P \approx 0.018$ and $i_P \approx 0.8^{\circ}$, while the capture into the Touma resonance that is nevertheless consistent with subsequent evolution requires pre-resonance $e=0.025$ and $i=0.25^{\circ}$. In our study of the next (i.e. previous in time) resonance we will fully consider both solutions and explore the conditions from which they could result.

\section{Yoder (1:2) Semi-Secular Solar Resonance}\label{sec:yoder}

\citet{yod82} was the first to realize that Phobos should have migrated through a semi-secular resonance with the Sun at 3.2~$R_M$. At this distance from Mars the orbital precession period of Phobos (both of longitude of pericenter $\varpi$ and the longitude of the ascending node $\Omega$) was two martian years. The main resonance with the Sun has the argument $2\varpi_P-\lambda_M-\varpi_M$, which we will refer as the Yoder resonance. Unlike the Touma resonance, which depended only on the orbit of Phobos and the obliquity of Mars, the strength of the Yoder resonance is directly proportional to the heliocentric eccentricity of Mars. 

\begin{figure}[ht]
\epsscale{1.15}
\plottwo{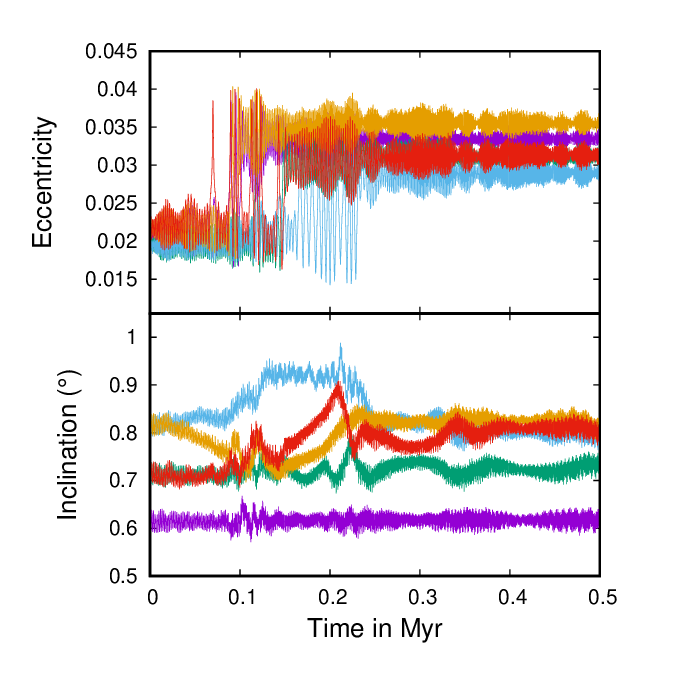}{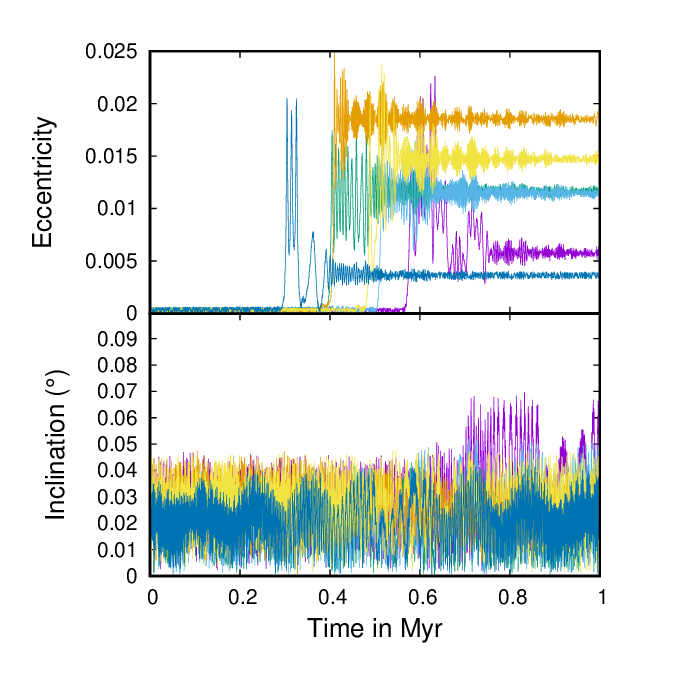}
\caption{Two sets of five different numerical simulations of Phobos's non-capture crossing through the Yoder (1:2) semi-secular resonance with the Sun at 3.2~$R_M$. In the left-hand panel, we started Phobos with substantial $e$ and $i$ with the post-resonance orbit bracketing the ``APH branch'' nominal orbital history. We can conclude that one set of possible pre-resonance orbital elements of Phobos is approximately $e_P = 0.02$ and $0.8^{\circ}$. In the right hand panel we started Phobos on a circular, planar orbit consistent with recent formation, with the outcome clearly inconsistent with the subsequent orbit of Phobos. Using standard tidal assumptions and ignoring any past ring torques, this crossing happened about 60~Myr ago.}
\label{yoder_kick}
\end{figure}

In the context of APH, \citet{yod82} has shown that this resonance gives Phobos only a modest ``kick'' in eccentricity with a negligible change in inclination. The left-hand side panels in Fig. \ref{yoder_kick} show simulations that include inward crossing of the Yoder resonance that are approximately consistent with our reconstructed history (i.e. post-resonance $e_P \approx 0.03$, $i_P \approx 0.8^{\circ}$). We confirm the prior result, and find that the increase of Phobos's eccentricity to $e_P \approx 0.03$ (as required by our model of tidal damping) requires a pre-resonance $e_P \approx 0.02$. 

However, if Phobos formed from a ring relatively recently, it should have formed at the fluid Roche limit which is approximately the same distance from Mars (3.2$R_M$) as the Yoder resonance \citep{hes17}. Due to all the modeling uncertainties in \citet{hes17}, it is not possible to say with confidence whether Phobos would form inside or outside the Yoder resonance. The Yoder resonance itself shifts outward in the presence of a ring. This accelerates the orbital precession of Phobos, but this is a negligible effect for ring masses comparable to Phobos'. Therefore, we must explore both possibilities of Phobos forming inside or outside the Yoder resonance. Right-hand panels of Fig. \ref{yoder_kick} show several simulations of Phobos crossing the Yoder resonance while migrating inward on a circular and planar orbit consistent with a very recent formation. While the eccentricity of Phobos sometimes receives kicks up to $e_P=0.02$, the inclination of Phobos is barely affected, making this scenario inconsistent with later evolution of Phobos's orbit. This result holds regardless of martian obliquity we used or the migration rate (the latter could have been slower than tidal migration alone due to presence of outward-acting ring torques).

\begin{figure}[ht]
\epsscale{1.15}
\plottwo{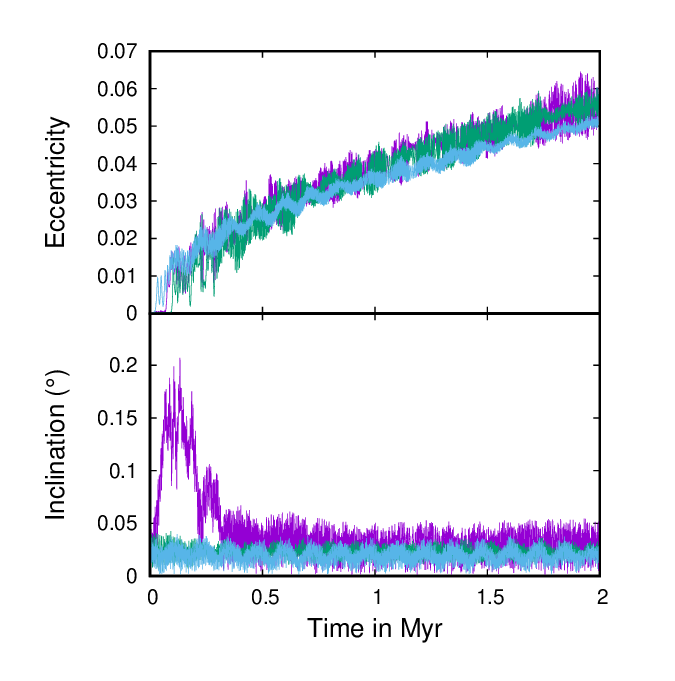}{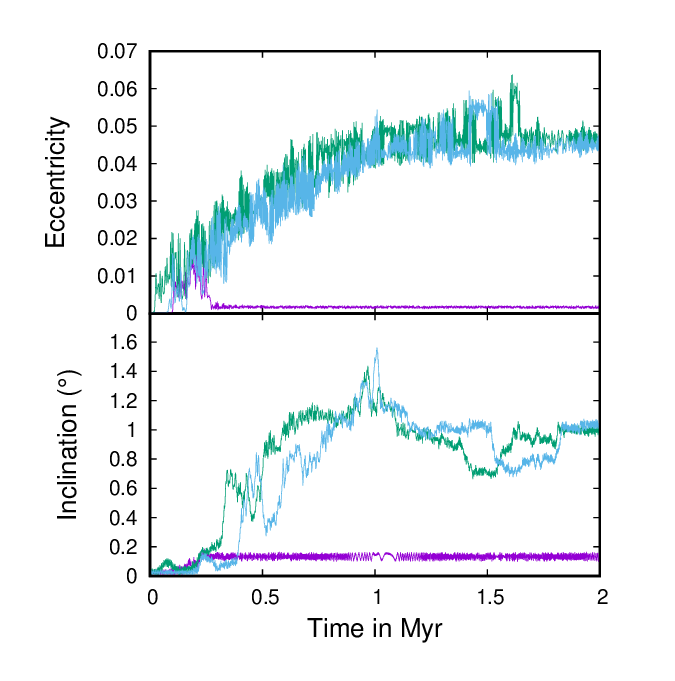}
\caption{Six different numerical simulations of Phobos's capture into of the Yoder (1:2) semi-secular resonance with the Sun at 3.2~$R_M$ when evolving outward due to disk torques. Phobos is assumed to be on a low-$e$, low-$i$ orbit immediately following formation from a ring (as suggested by RPH). The three simulations on the left-hand side used the initial axial tilt of Mars of $25^{\circ}$ (light blue), $35^{\circ}$ (green) and $40^{\circ}$ (purple), while the three simulation on the right used the initial axial tilt of Mars of $45^{\circ}$ (green), $50^{\circ}$ (purple) and $55^{\circ}$ (light blue).}
\label{yoder_capture}
\end{figure}

The other possibility is that Phobos formed from the outer edge of the disk just interior to the Yoder resonance, and then migrated outward into the resonance due to ring torques. Outward migration by a satellite is generally conducive to capture into eccentricity-type semi-secular resonances such as evection \citep{tou98} and we expect this to be true of the Yoder resonance. Figure \ref{yoder_capture} shows six different simulations of the capture of outward-migrating Phobos into the Yoder resonance. In these simulations we allowed the eccentricity of Mars to realistically vary due to planetary perturbations, but these variations had a relatively negligible effect on the stability of the resonance. Note that our planetary initial conditions were those of the present epoch, and that the Yoder resonance may behave differently at times when the eccentricity of Mars was much lower. Unlike martian obliquity, which does not have significant back-reaction on the planetary orbits and can be adjusted with no side-effects, modification of martian orbital eccentricity would have to be done as a part of modeling the possible past evolution of the Solar System's secular eigenmodes, which we leave for future work.

In simulations in which the obliquity of Mars was low (25$^{\circ}$-35$^{\circ}$) (left-hand side of Fig. \ref{yoder_capture}) Phobos' eccentricity steadily increases in the resonance (keeping its rate of orbital precession constant) while the inclination is not affected at all (as expected as the Yoder resonance does not involve inclination). However, integrations with martian obliquity of $\theta_M > 45^{\circ}$ (right-hand side of Fig. \ref{yoder_capture}) the inclination of Phobos is highly chaotic and can reach $i_P > 1^{\circ}$ in a few Myr. We believe that the source of this dynamical chaos is the overlap between the Yoder resonance (with argument $2\varpi_P-\lambda_M-\varpi_M$) and another, weaker, solar semi-secular resonance involving inclination with argument $2\varpi_P-2\lambda_M-2 \Omega_P+2 \Omega_{Eq}$. 

The difference between the Yoder argument and the half of the inclination resonance's argument is $\Psi_Y=\varpi_P+\Omega_P-\varpi_M-\Omega_{Eq}$. This may appear like an argument of a secular resonance, but this is accidental, as only semi-secular resonances appear to significantly affect Phobos. When $\Psi_Y$ is slow moving or stationary, this simply indicates that the two semi-secular resonances are overlapping and producing dynamical chaos. In particular, the rate of change of $\Omega_{eq}$ (the longitude of the martian vernal equinox) depends directly on the obliquity of Mars, making it possible for $\Psi_Y$ to be stationary and chaos to arise for some values of martian obliquity. Note that the angle sum $\varpi_P+\Omega_P$ involves two angles that precess rapidly in opposite directions with almost equal rates, so their sum has circulation periods comparable to those of planetary secular modes. 

Simulations plotted in Figure \ref{yoder_capture} serve as a proof of concept that the outward migration into Yoder resonance can generate eccentricities and inclinations sufficient to explain subsequent dynamical evolution of Phobos down to the present. However, there are many uncertainties about the interaction of Phobos with both the Yoder resonance and the martian ring. As the ring is losing mass to infall onto Mars, Phobos's migration would be dominated more and more by tidal dissipation rather than ring torques. 

It is possible that Phobos could evolve outward in the Yoder resonance, stall and then migrate back inward. However, this scenario would be inconsistent with later constraints, as the reverse migration within the resonance would reduce Phobos's eccentricity to $e_P< 0.01$, at which point the Yoder resonance breaks. Therefore, in order for the RPH hypothesis to hold, Phobos would need to fully migrate outward through the Yoder resonance. 

\begin{figure}[ht]
\epsscale{1.15}
\plottwo{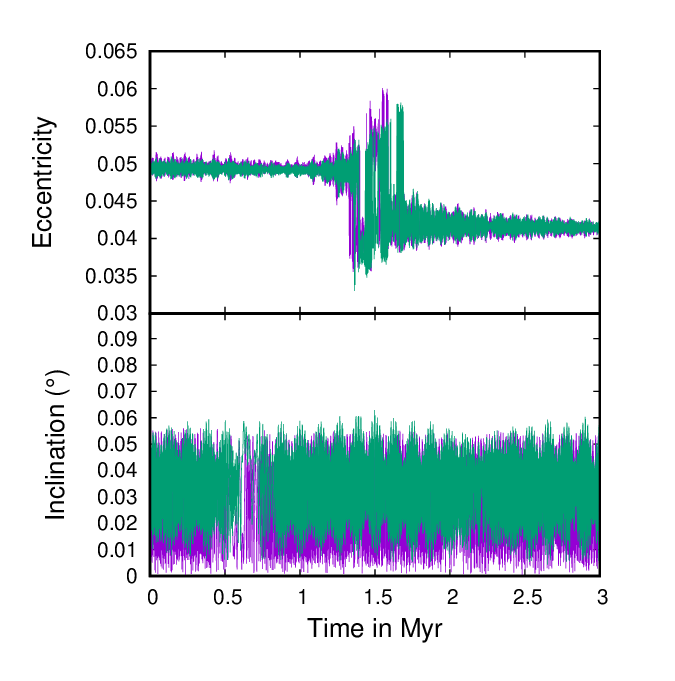}{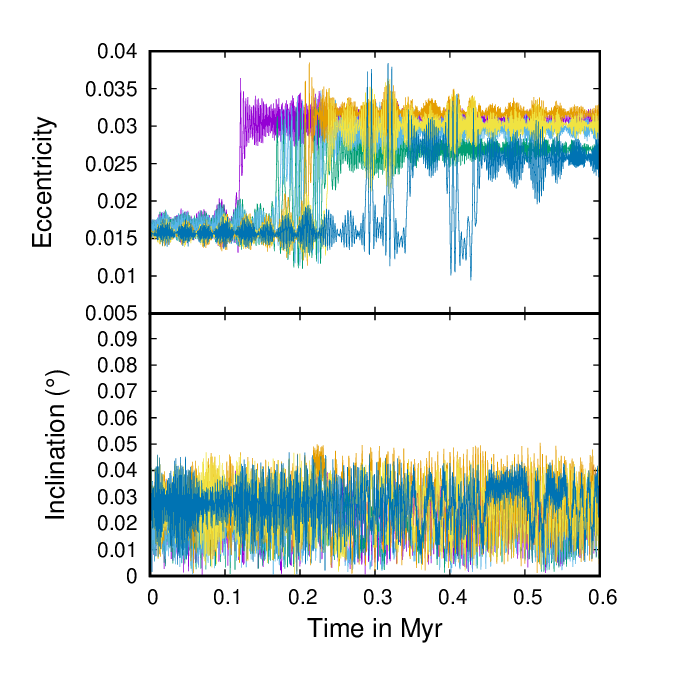}
\caption{Two sets of numerical simulations of Phobos's crossing of the Yoder (1:2) semi-secular resonance with the Sun at 3.2~$R_M$ with high initial eccentricity and near-zero inclination. The purpose of these simulations is model the evolution of Phobos after the end of capture in the Yoder resonance, with the focus on potential separate excitation of Phobos's inclination. The left-hand panel shows the crossing of the resonance while Phobos is moving outward due to ring torques, while the right-hand panels shows the inward crossing of the resonance by Phobos once tides become dominant. The two simulations on the left-hand side used the initial axial tilt of Mars of $25^{\circ}$ and $45^{\circ}$, while the six simulation on the right used these two obliquities and three different net migration rates (equivalent to martian $Q=80, 100, 120$).}
\label{post_capture}
\end{figure}

We also find some indications that the Yoder resonance may break spontaneously in some of the more chaotic cases, but we have not observed this for cases when Phobos is deep within the resonance with high eccentricity. The most likely outcome, in our opinion, is that Phobos physically collides with the outer ring edge at pericenter, which would eject it from the resonance with slightly smaller $a$ and $e$ (but the same pericenter distance). In the ring-collision scenario Phobos would then migrate outward again through the Yoder resonance, this time with substantial $e$ (and possibly also $i$). And in all scenarios in which the Yoder resonance was broken and Phobos migrated past it also require Phobos to cross the Yoder resonance one last time when migrating inward. 

In Fig. \ref{post_capture} we plot some of the simulations of both outward and inward crossings of the Yoder resonance by an already eccentric Phobos. In these simulations we chose to keep inclination low to determine whether the observed inclination of Phobos could have been generated in these later crossings. Overall, we find that outward crossing reduces $e$ and inward crossing increases $e$, with no effects to inclination, consistent with results of Fig. \ref{yoder_kick}. We conclude that the subsequent passages through the Yoder resonance by young Phobos after the resonance capture is over do not fundamentally change the dynamical picture. The $e$ and $i$ of Phobos in the RPH would need to be largely generated during the capture in the Yoder resonance, which requires outward migration of Phobos after its formation interior to this resonance at 3.2~$R_M$.

Our study of the Yoder resonance has demonstrated that the RPH is in principle a viable scenario for the origin of Phobos from the point of view of its current orbital elements. In order to acquire enough eccentricity and inclination,  as long as Phobos formed interior to the Yoder resonance at 3.2~$R_M$ and migrated outward into this resonance at rates slower than the tidal migration rate (i.e. we assume the absolute value of the ring torque migration was less than two times that of the tidal evolution). In the context of APH, we find that Phobos must already have $e_P \approx 0.02$ and $i_P \approx 0.8^{\circ}$ when encountering the Yoder resonance during its inward tidal migration. Therefore, earlier resonances, such as the 2:1 tesseral resonance at 3.8~$R_M$ (addressed in Section \ref{sec:2to1}) would need to be responsible for most of the $e$ and $i$ of Phobos if it is ancient. Additionally, an ancient Phobos will need to be reconciled with the low current eccentricity of Deimos, as pointed out by \citet{yod82}. We address the constraints on the evolution of Phobos stemming from the dynamics of Deimos in the next section.

\section{Phobos-Deimos 3:1 and 2:1 MMRs and the Long-Term Dynamics of Deimos}\label{sec:deimos}

While \citet{yod82} generally found that the orbit of Phobos is consistent with it forming closer to the synchronous orbit and migrating inward over the age of the Solar System, he did find one inconsistency in that scenario. Phobos is expected to have migrated through the 2:1~MMR with Deimos at 4.3~$R_M$, and during that resonance crossing Deimos is expected to acquire $e_D=0.002$, which is an order of magnitude above Deimos's current eccentricity. Multiple authors have proposed that Deimos may be be more tidally dissipative than expected~\citep{bra20, qui20, bag21}. However, \citet{yod82} persuasively argued that if one makes the reasonable assumption that both Phobos and Deimos have similar tidal properties to each other, then Phobos would damp its eccentricity too rapidly to be consistent with observational constraints. Thus, enhanced tidal dissipation of Deimos is a highly unlikely solution for its low eccentricity. 

An obvious alternative explanation for lack of excitation of Deimos's eccentricity is that Phobos never crossed the 2:1 MMR with Deimos because it formed closer to Mars. While this is not plausible in the APH \citep{yod82}, it could be used as an argument for the RPH, as in that model Phobos formed at about 3.2~$R_M$ and did not migrate outward far enough to reach the 2:1 MMR~\citep{hes17}. We can extend this argument to other MMRs between Phobos and Deimos, such as the 3:1 MMR at 3.33~$R_M$.

\begin{figure}[ht]
\epsscale{.6}
\plotone{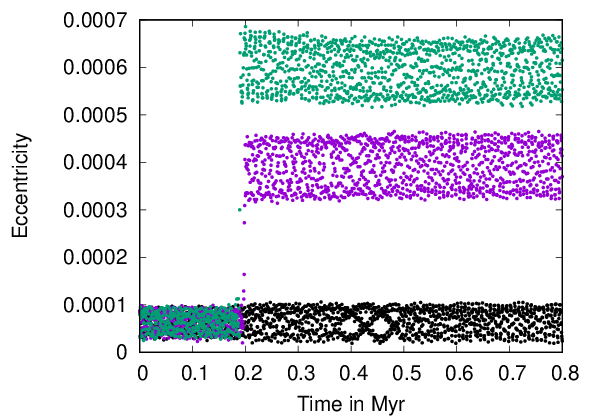}
\caption{Eccentricity of Deimos during the crossing of 3:1 Phobos-Deimos MMR as Phobos was migrating inward. If we assume that Phobos had very low eccentricity, there is little effect on Deimos (black points), but if the eccentricity of Phobos was $e_P=0.01$ (magenta) or $e_P=0.02$ (green) at the time, Deimos's eccentricity is excited above the observed value of $e_D = 0.00027$. Using standard assumptions this crossing happened about 100 Myr ago.}
\label{31mmr}
\end{figure}

Fig. \ref{31mmr} shows three simulations of divergent crossing of the 3:1 Phobos-Deimos MMR, assuming that Phobos is migrating inward at our standard rate (i.e. martian $Q=80$, $k_2=0.14$). The there sets of points assume a zero-eccentricity Phobos (black), $e_P=0.01$ (magenta) and $e_P=0.02$ (green). For eccentric Phobos the resulting eccentricity of Deimos is in both cases larger than the observed $e_D=0.00027$ \citep{jac14}. The eccentricity of Phobos is relevant here because this is a second order MMR, with the strongest eccentricity argument being $3\lambda_D-\lambda_P-\varpi_P-\varpi_D$, so the corresponding Hamiltonian term will be proportional to $e_P$. Our findings from the previous section suggest a $e_P \simeq 0.02$ at 3.2~$R_M$, which combined with our estimate of eccentricity damping gives us $e_P \simeq 0.026$ at 3.33~$R_M$. Given the results shown in Fig. \ref{31mmr}, this $e_P$ directly implies that the Phobos-Deimos 3:1~MMR crossing may be inconsistent with observed $e_D$. 

\begin{figure}[ht]
\epsscale{.6}
\plotone{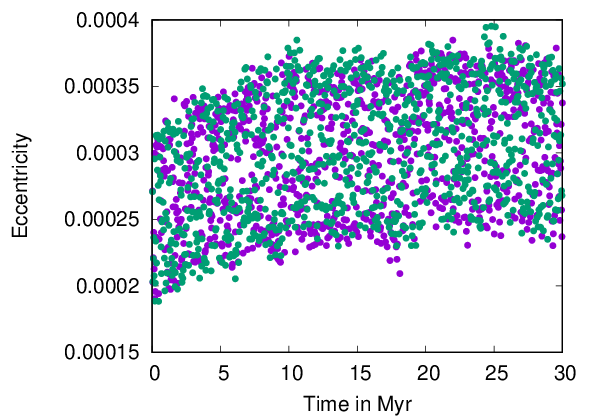}
\caption{Eccentricity of Deimos in two simulations of long-term evolution of Deimos's orbit that start from the present configuration. Green points have been obtained using numerical integrator {\sc s-simpl}, which fully integrated the dynamics of Deimos, eight planets and the martian spin axis. The purple points were obtained using {\sc s-simpl-read} integrator, in which Deimos's dynamics is simulated using pre-calculated positions for the planets and the martian spin axis.}
\label{ssimpl2}
\end{figure}

Looking at the bigger picture beyond the 3:1~MMR, we can firmly conclude that Deimos's current eccentricity appears to be too low. Apart from the 2:1~MMR crossing noted by \citet{yod82}, ancient resonances with extinct moons should have excited Deimos's eccentricity to values comparable to the present one \citep{cuk20}, and jumps $e_D$ shown in Fig. \ref{31mmr} are cumulative, meaning that already eccentric Deimos will become more eccentric, rather than experiencing a random walk in eccentricity. Therefore we decided to follow a speculative suggestion by \citet{yod82} that there might be unknown effects on the orbit of Deimos and examine the dynamics of Deimos in some detail using full numerical simulations.

The green points in Fig. \ref{ssimpl2} show the eccentricity of Deimos in a 30~Myr simulation that includes the entire planetary system and variable obliquity of Mars. The initial conditions were based on the current state of the system. It is clear that Deimos's eccentricity is affected by perturbations that act on timescales longer than 30~Myr and may even be chaotic in the long-term. On the other hand, the inclination of Deimos appears to be much less affected by these long term processes than the eccentricity. We find that the most plausible source of these perturbations are secular resonances with the planets that involve the argument $\varpi_D+\Omega_D$, somewhat like the secular resonance of Iapetus found by \citet{cuk18}. 

While the periods of circulation of Deimos's longitudes of pericenter and ascending node (the latter measured w.r.t the martian equator) are on the order of fifty years and much faster than secular periods of the planetary system, these precessional motions are almost exactly equal in period and opposite in direction so that their sum circulates very slowly, with a period on the order of $10^5$~yr. Such long periods put the $\varpi+\omega$ precession in the right frequency range to be coupled to the planetary secular modes, explaining the very longer period (and possibly chaotic) perturbations. Furthermore, if Deimos is affected through the $\varpi_D+\Omega_D$ argument, that means that the perturbations of eccentricity and inclination (in radians) of Deimos should be fully correlated and also independent of the values of $e_D$ and $i_D$. Therefore, while the free inclination of Deimos may also experience variations on the order of $10^{-4}$~rad, this is equivalent to 
0.005$^{\circ}$ and therefore not noticeable relative to the free eccentricity of almost two degrees.

\begin{figure}[ht]
\epsscale{.6}
\plotone{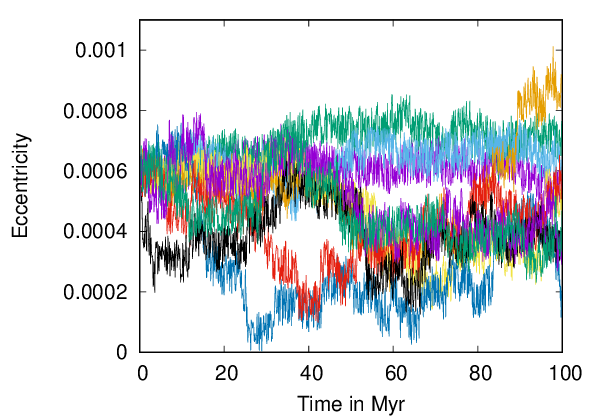}
\caption{Eccentricity of Deimos in a set of ten simulations of long-term evolution of Deimos's orbit that were completed using {\sc s-simpl-read} integrator. Each simulation was started with $e_D=0.0006$ and $i_D=1.8^{\circ}$, but with different axial tilts of Mars ranging from $22^{\circ}$ and $40^{\circ}$. The evolution of the planetary orbits in all of these simulations is the same and starts from the current configuration.}
\label{lsread_all}
\end{figure}

The major implication of Fig. \ref{ssimpl2} is that if the eccentricity of Deimos is not stable over long periods of time, then we cannot rely on it to constrain the past resonances between Phobos and Deimos. To fully explore this question we need to be able to integrate the orbit of Deimos with all the relevant perturbations included for sufficiently long periods of time. Additionally, the chaotic nature of Mars's orbital and spin axis dynamics requires us to explore a wider phase space, particularly with respect to martian obliquity. Unfortunately, green points in Fig. \ref{ssimpl2} that were produced using {\sc s-simpl} (Table \ref{integrators}) took an excessive amount of CPU time. The reason for that is that the timestep is dictated by the orbital period of Deimos, which slightly longer than a day, so {\sc ssimpl} was forced to use the same timestep when integrating the planets. 

In order to overcome the timestep problem, we modified {\sc s-simpl} into a version called {\sc s-simpl-read} which reads the parameters of martian orbit and spin axis orientation from the results of an earlier integration. The initial integration does not include Deimos, and therefore could be done with orders of magnitude longer timestep (which is now dictated by the orbital period of Mercury). The second integration interpolates the position of Mars relative to the Sun and the orientation of the martian spin axis from the results of the first, and uses those values to implement the perturbations on Deimos that arise from the martian oblateness and the Sun. Here we ignore the direct perturbations on Deimos from the planets, but we were able to test their importance directly using {\sc s-simpl} (which has a ``switch'' that can turn on or off direct perturbations on satellites of one planet by other planets in the system) and found them to be unimportant. The purple points in Fig. \ref{ssimpl2} are produced using {\sc s-simpl-read}, which is about three times faster than {\sc s-simpl}; the overall good correspondence to major trends found using {\sc s-simpl} is encouraging. 

\begin{figure}[ht]
\epsscale{.6}
\plotone{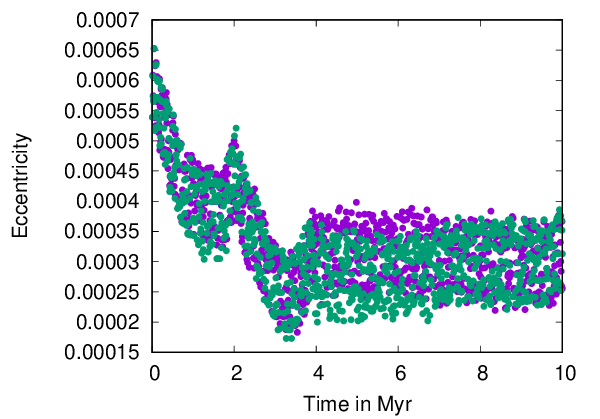}
\caption{Eccentricity of Deimos in two 10 Myr simulations of Deimos's orbit that were conducted in order to compare numerical integrators. The purple points plot the beginning one of the {\sc s-simpl-read} integrations from Fig. \ref{lsread_all} (with $\epsilon=34^{\circ}$); this integration showed most drastic changes in $e_D$ within the first few Myr. The green points plot an integration with same initial conditions done using {\sc s-simpl} (i.e. with Deimos and the planets integrated directly and together, and not in two stages). The overall similarity of $e_D$ changes confirms that {\sc s-simpl-read} is able to capture the most relevant sources of Deimos's chaotic dynamics despite being computationally less expensive.}
\label{ssimpl7}
\end{figure}

Figure \ref{lsread_all} plots ten different 100~Myr simulation of Deimos's orbit done using {\sc s-simpl-read}. Unlike in the Fig. \ref{ssimpl2}, we use initial $e_D=0.0006$, inspired by the outcome of 3:1 Phobos-Deimos MMR in Fig. \ref{31mmr}. The initial axial tilt of Mars was varied from 22$^{\circ}$ to 40$^{\circ}$, while the free inclination of Deimos was kept at $1.8^{\circ}$. Fig. \ref{lsread_all} shows that Deimos experiences significant and chaotic variations in eccentricity over 100~Myr. This conclusively demonstrates that any excitation of Deimos's eccentricity by the 3:1 MMR with Phobos could be erased by chaotic variation, and that we have no strong constraints on the past eccentricity of Deimos (assuming eccentricities $e_D << 0.01$). The Phobos-Deimos 2:1 MMR explored by \citet{yod82} may have produced $e_D=0.002$, which is factor of three above the initial value in Fig. \ref{lsread_all}, but also likely happened an order of magnitude earlier ($\simeq 1$~Gyr), giving the chaotic evolution more time to act and possibly reduce Deimos's eccentricity. 

\begin{figure}[ht]
\epsscale{.6}
\plotone{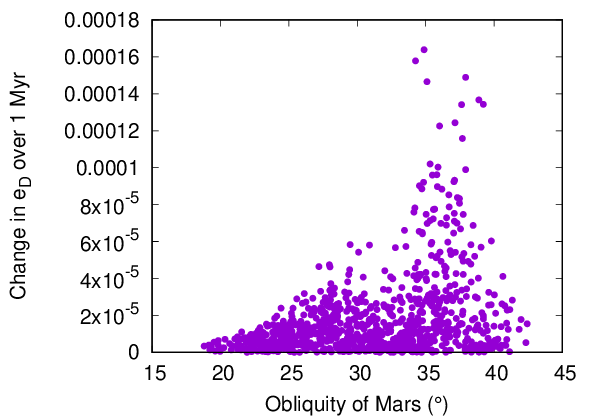}
\caption{Dependence of eccentricity change of Deimos over 1~Myr on the obliquity of Mars. Each of the ten simulations in Fig. \ref{lsread_all} have been divided into 100~1-Myr segments, and the averaged $e_D$ and obliquity of Mars (w.r.t. its heliocentric orbit) $\theta$ were calculated for each segment. Individual points plot magnitude of the difference between average $e_D$ of each segment and its predecessor as a function of average $\theta$ for the segment. There appear to be  secular resonances affecting the orbit of Deimos when the obliquity of Mars is in the $33^{\circ}-38^{\circ}$ range.}
\label{delta_e}
\end{figure}

Figure \ref{ssimpl7} illustrates the comparison between a direct integration and the {\sc s-simpl-read} two-step scheme. The simulation in Fig. \ref{lsread_all} that started with the martian axial tilt $\epsilon=34^{\circ}$ experienced very rapid decline in eccentricity of Deimos (purple points), and we wanted to check that this behavior is real. We re-integrated the first 10~Myr of this run using {\sc s-simpl} and obtained very similar results (green points) that also show a dramatic drop in $e_D$. We consider this to be a direct proof that {\sc s-simpl-read} captures the most relevant features of the long-term dynamics of Deimos.

Figure \ref{delta_e} plots changes of mean $e_D$ over 1 Myr as a function of martian obliquity. The variability of Deimos's eccentricity at low martian obliquities (cf. Fig. \ref{ssimpl2}) is lower than at obliquities above $\theta=33^{\circ}$. We tentatively propose that there may be secular resonances experienced by Deimos when the martian obliquity is in the $33^{\circ}<\theta<38^{\circ}$ range. We were unable to identify individual long-term secular resonances between Deimos and the planets, such as the one found by \citet{cuk18} for Iapetus. We find that small changes in the orbit of Deimos produce significant variation in the precession rate of $\varpi_D+\Omega_D$, making any secular resonance arguments chaotic and librating only intermittently. 

We hope that future work will be able to better understand these secular interactions, and resolve whether larger apparent excursions in $e_D$ to lower values in Fig. \ref{lsread_all} are systematic or stochastic. In order to study the realistic history of Deimos rather than its general dynamics, it will be necessary to introduce precise reconstructions of martian past obliquity and orbit \citep[e.g. ][]{zee22} into the simulations of Deimos's orbit. This paper, however, is focused on the  dynamical history of Phobos. For the purposes of this paper, we can conclude that the eccentricity of Deimos is not stable over the long-term and therefore cannot be used to constrain the past evolution of Phobos. 

\bigskip

\section{2:1 Tesseral Resonance and Associated Three-Body Resonances}\label{sec:2to1}

\citet{yod82} found that the 2:1 tesseral resonance between the orbital motion of Phobos and martian rotation was a single most significant perturbation to the orbit of Phobos over its history. In the context of the APH, this resonance is expected to be the main source of Phobos's eccentricity and inclination, which were subsequently only somewhat modified by the Yoder, Touma and 3:1 tesseral resonances. \citet{yod82} found that the main terms in the Hamiltonian should produce kicks of approximately $\Delta e=0.04$ and $\Delta i=0.4^{\circ}$ to Phobos's eccentricity and inclination, respectively, as Phobos migrates inward through the 2:1 tesseral resonance. Note that there can be no capture into any of the tesseral resonances during tidal evolution, as tides make the moons migrate away from the synchronous orbit, while capture into tesseral resonances is only possible for migration toward the synchronous orbit. More recently, \citet{qui20} have explored the 2:1 tesseral crossing using a purely numerical simulation (accelerated 1000 times over expected tidal migration rate) and found kicks in eccentricity and inclination of Phobos that closely match the predictions of \citet{yod82}.

\begin{figure}[ht]
\epsscale{.6}
\plotone{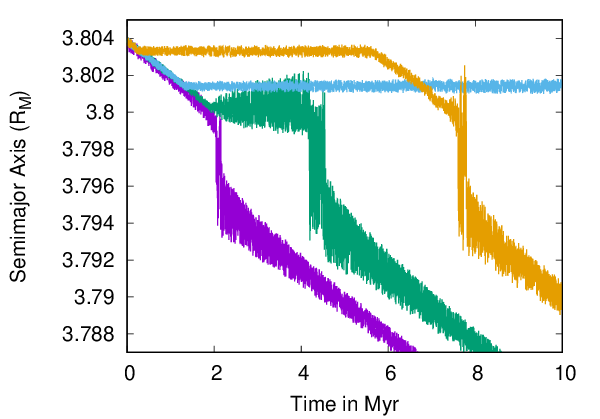}
\caption{Four simulations of Phobos encountering the 2:1 tesseral resonance with the martian rotation. Gold and light blue lines show simulations with zero initial inclination of Phobos, and these simulations show captures into different inclination-type tesseral three-body resonances involving Phobos, martian figure and the Sun. The green line shows a simulation with a non-zero inclination and low eccentricity which experiences a temporary capture into the principal eccentricity-type tesseral three body resonance. The purple line shows a simulation with Phobos starting both inclined and eccentric in which Phobos crosses the 2:1 tesseral resonance without any captures into tesseral three-body resonances. Under our assumptions this crossing happened 300~Myr ago.}
\label{isphob_a}
\end{figure}

Our own simulations of a highly accelerated evolution of Phobos match past results of \citet{yod82} and \citet{qui20}. However, simulations using realistic tidal migration rates produce dramatically different results, including resonance capture, i.e. the inward migration of Phobos being stopped near the 2:1 tesseral resonance. Figure \ref{isphob_a} illustrates several ways this capture may happen, where, depending of the initial eccentricity and inclination of Phobos, it can get captured in several different sub-resonances that have slightly different semimajor axes. When we analyzed these unexpected results, we realized that the relevant sub-resonances that lead to capture are not the main tesseral terms, but involve a combination of perturbations from the martian figure and the Sun. This interaction between the martian gravity harmonics, the Sun and Phobos is in many ways similar to three-body resonances, so we will refer to these resonances as tesseral three-body resonances (T3BR). 

While the capture into two-body resonances requires convergent orbital evolution \citep{md99}, diverging orbits can be captured into T3BRs as the presence of the third body makes conservation of energy and angular momentum more complex and allows for new types of behavior. The strongest T3BR (green line in Fig. \ref{isphob_a}) has an argument $2\phi_M - \lambda_P - 2 \lambda_M + \varpi_P$ (where $\phi_M$ is the sidereal rotation angle of Mars) and makes the eccentricity of Phobos grow while the semimajor axis stays constant. Another type of T3BR capture plotted in Fig. \ref{isphob_a} by a light blue line is into an inclination-type argument $4\phi_M-2\lambda_P - 4\lambda_M + 2 \Omega_P$, which leads to increasing inclination. Higher-order T3BRs are also observed, such as the inclination-type plotted by the gold line in Fig. \ref{isphob_a}. However, these higher-order T3BRs depend on the eccentricity and inclination of Mars itself which are variable, leading to capture into higher-order $i$-type T3BR being temporary (as in Fig. \ref{isphob_a}), and capture into the higher-order $e$-type T3BR being impossible once the full dynamics of martian eccentricity is included. 

\begin{figure}[ht]
\epsscale{.6}
\plotone{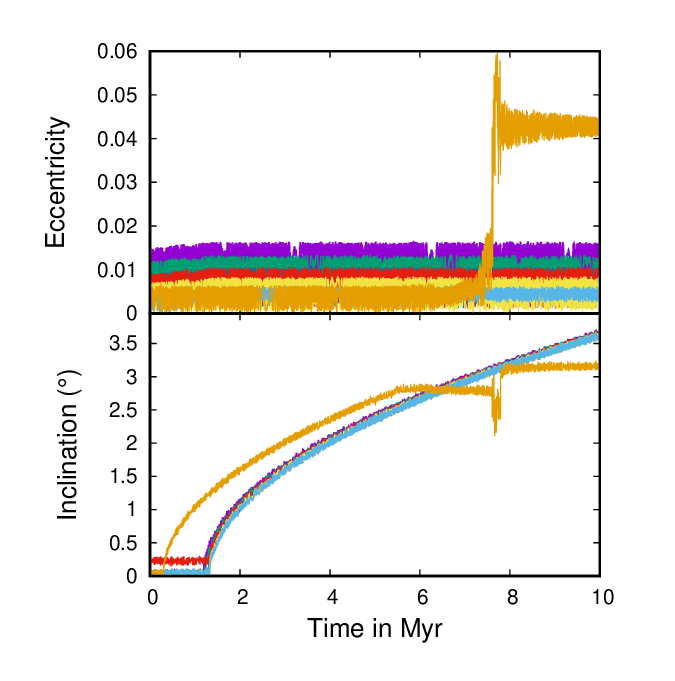}
\caption{Seven simulations of Phobos encountering the 2:1 tesseral resonance with martian rotation resulting in capture into the inclination-type tesseral three-body resonances involving Phobos, martian figure and the Sun. The gold line (same simulation as the gold line in Fig. \ref{isphob_a}) plots a rare capture into a higher-order inclination-type tesseral thre-body resonance. While in all other runs, capture into the lowest-order inclination-type tesseral three-body resonance persists to the end of the simulation, the higher-order capture is temporary.}
\label{isphob_ires}
\end{figure}

We divide the outcomes of the encounter of Phobos into three classes: $i$-type T3BRs, $e$-type T3BRs, and no-capture resonance crossings (e.g. purple line in Fig. \ref{isphob_a}). To show the range of relevant dynamical behavior, we use a grid of 18 simulations, which were started near the 2:1 tesseral resonance with inclinations of $i_P=0, 0.1^{\circ}, 0.2^{\circ}$ and six eccentricities in the $e_P=0-0.01$ range (which results in free eccentricities $e_{P,free}=0.002-0.012$; forced eccentricity is due to the main 2:1 tesseral resonance). All initially planar orbit simulations and only one inclined simulation (with $i_P=0.2^{\circ}$) resulted in $i$-type T3BR captures, plotted in Fig. \ref{isphob_ires}. 

One of the seven simulations featured capture into a higher-order $i$-type T3BR, while others were captured into the main $i$-type T3BR with the argument $4\phi_M-2\lambda_P+4\lambda_M-2\Omega_M$. The higher-order capture was temporary, while the main $i$-type T3BR captures persistent to the end of the simulation, reaching inclinations of almost 4 degrees. The temporary capture leaves Phobos with $i_P > 3^{\circ}$, which is too high compared to the current value. Furthermore, continued growth of inclination during T3BR capture can lead to a sesquinary catastrophe \citep[][see Section \ref{sec:intro}]{cuk23}, which would result in Phobos re-accreting on a circular and planar orbit interior to the 2:1 tesseral resonance. 

As we found in the previous section, formation on a circular, planar orbit between 3.2 and 3.8~$R_M$ is one of the scenarios we can rule out for Phobos. Therefore, we tentatively conclude that the capture of Phobos into the $i$-type T3BR adjacent to the 2:1 tesseral resonance did not happen. This constraint implies that the inclination of Phobos preceding the 2:1 tesseral resonance had to be non-zero. We can confirm that $i_P=0.1^{\circ}$ is sufficient to avoid the resonance capture, and more work is needed to identify the exact critical inclination for this process.

\begin{figure}[ht]
\epsscale{1.15}
\plottwo{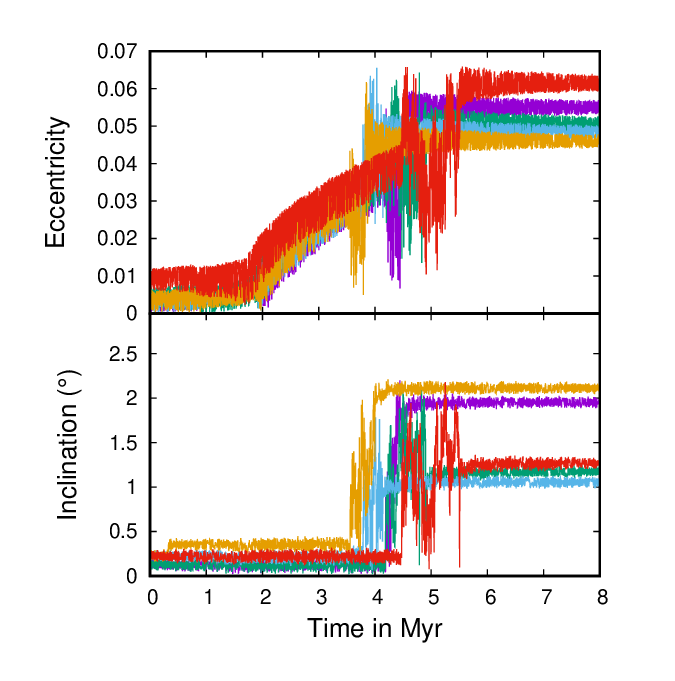}{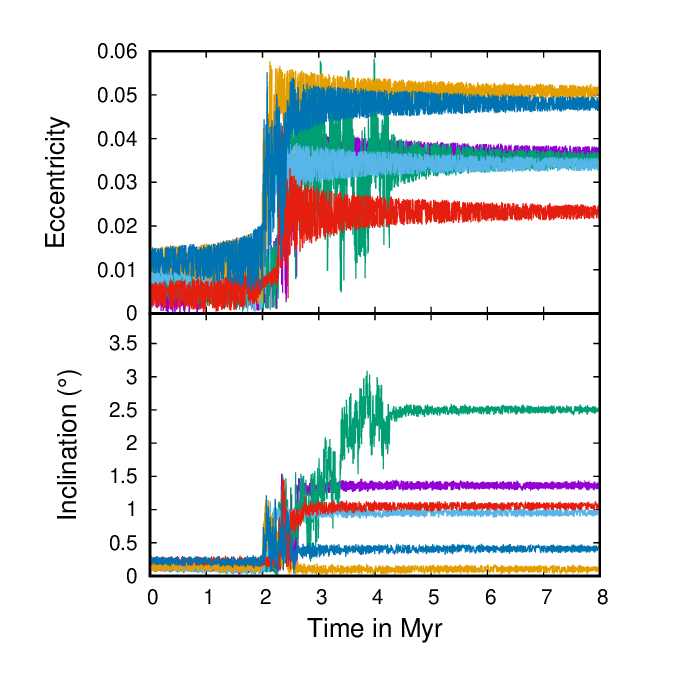}
\caption{Left-hand panel: five simulations of Phobos encountering the 2:1 tesseral resonance with martian rotation resulting in temporary capture into an eccentricity-type tesseral three-body resonance involving Phobos, the martian figure, and the Sun. Right-hand panel: Six simulations of Phobos encountering the 2:1 tesseral resonance with martian rotation that experienced no resonant capture.}
\label{isphob_eres}
\end{figure}

The left-hand panel in Fig. \ref{isphob_eres} shows the simulations in which Phobos was captured into $e$-type T3BRs. In all cases this capture was temporary, and Phobos exited the T3BR once its eccentricity reached about $e_P=0.04$. The reason for breaking of the T3BR is illustrated by the green line in Fig. \ref{isphob_a}: as Phobos becomes more eccentric within the T3BR, its semimajor axis exhibits wider variations and eventually starts overlapping with the main 2:1 tesseral resonance, triggering dynamical chaos. After the T3BR exit, Phobos typically experiences a period of chaos within the 2:1 tesseral resonance, after which it exits the resonance with substantial eccentricity and inclination. 

We can compare these outcomes with the simulations that did not experience T3BR capture (right-hand panel in Fig. \ref{isphob_eres}). Some of the no-capture simulations show a short-lived chaotic residence in the 2:1 tesseral resonance, followed by resonance exit. Overall, simulations with higher initial eccentricities were less likely to experience $e$-type T3BR capture, but we did observe such capture for one of the simulations with $e_{P,free}=0.01$. The final eccentricities and inclinations for both capture and non-capture simulations were distributed over a wide range of outcomes, with simulations featuring $e$-type T3BR capture generally having higher post-resonance eccentricities.  

The 2:1 tesseral resonance is only relevant in the context of APH, as the RPH predicts that Phobos formed well interior to this resonance. The range of $e_P$ and $i_P$ resulting from the 2:1 tesseral resonance crossing (Fig. \ref{isphob_eres}) is wholly consistent with tidal evolution to through the subsequent resonances to the present state. A no-capture crossing through the Touma resonance would require $e_P \simeq 0.02$ and $i_P \simeq 0.75$ immediately before the Yoder resonance (Fig. \ref{yoder_kick}). Assuming our adopted estimate of dissipation within Phobos resulting in $e_P \propto a_P^{6.46}$ \citep{yod82}, these constraints would imply $e_P \simeq 0.06$ after the 2:1 tesseral resonance. This is comparable to the eccentricities in Fig. \ref{isphob_eres}, possibly indicating that Phobos may have been captured in an $e$-type T3BR, assuming that the estimate of dissipation within Phobos we used is correct. 

We note that no outcomes involving the $e$-type T3BR have final $i_P<1^{\circ}$; however we feel that the limited number of (numerically expensive) simulations that we have completed so far do not allow us to exclude this evolutionary pathway on the basis of resulting inclinations. The only firm conclusions we can draw from this Section is that the present orbit of Phobos is consistent with it crossing the 2:1 tesseral resonance, with the constraint that pre-resonance needs to be $i_P>0.1^{\circ}$ in order to avoid capture into the $i$-type T3BR.

\section{3:2, 4:3 and 5:4 Tesseral Resonances}\label{sec:early}

\citet{yod82} concluded that the orbit of Phobos is consistent with formation on a circular and planar orbit outside the 2:1 tesseral resonance and did not explore more distant tesseral resonances. In the previous section we found that Phobos must have had at least a small inclination prior to encountering 2:1 tesseral resonance in order to avoid capture into the $i$-type T3BR. The capture into an $i$-type T3BR would be incompatible with the present state, as the resulting $i_P$ would be either too high or too low (the latter if Phobos re-accreted after a sesquinary catastrophe). Therefore, we need another dynamical mechanism that acted on Phobos before its 2:1 tesseral resonance crossing to excite $i_P$ by at least $0.1^{\circ}$. As our nominal tidal assumptions put the crossing of the 2:1 tesseral resonance few hundred Myr ago, in the context of APH it is very likely that Phobos crossed additional tesseral resonances closer to the synchronous orbit.

\begin{figure}[ht]
\epsscale{1.15}
\plottwo{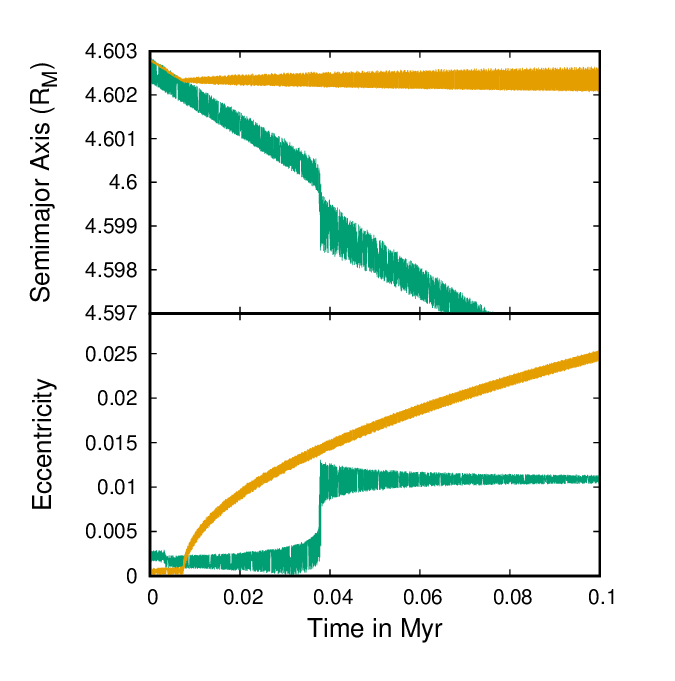}{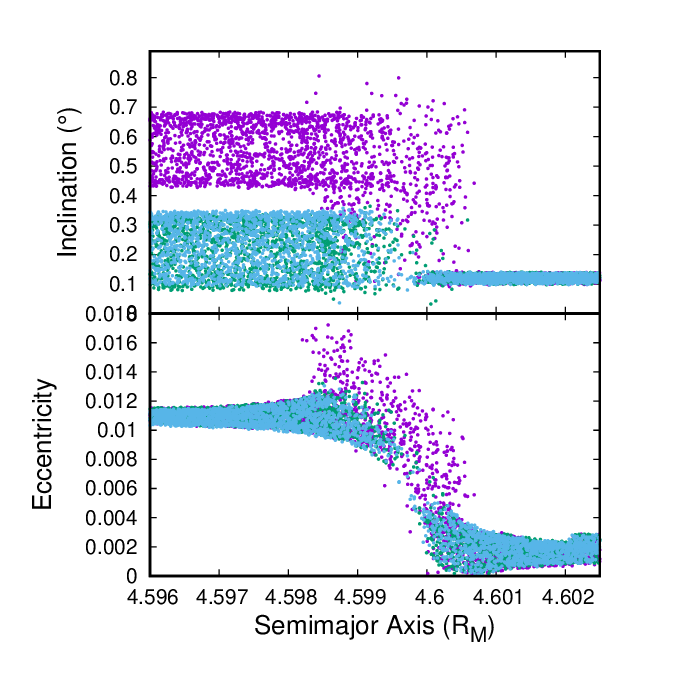}
\caption{Left-hand panel: Two 100x accelerated simulations of Phobos encountering the 3:2 tesseral resonance. The simulation with low initial eccentricity (gold) results in capture into the eccentricity-type tesseral three-body resonances involving Phobos, martian figure and the Sun. while the simulation with initial $e_P=0.002$ (green) crosses the resonance without capture. Right-hand side: Three simulations of Phobos crossing the 3:2 tesseral resonance without a resonance capture. The purple points show the simulation with the realistic migration rate, while the blue and green points show runs with Phobos's migration accelerated 10 and 100 times, respectively. The inclination is plotted relative to the martian equator and includes a forced inclination of $0.12^{\circ}$. Assuming constant martian $Q$, this resonance happened about 1.1~Gyr ago.}
\label{early}
\end{figure}

When exploring all possible tesseral resonances between Phobos and the martian rotation, we found that we need not consider resonances beyond the first order, with the exception of 3:1 tesseral resonance (Section \ref{sec:31res}). Tesseral resonances in some ways resemble resonances with an imaginary satellite that is at martian synchronous orbit, which helps explain why capture into them is impossible during a satellite's tidal evolution away from the synchronous radius. However, the source of the perturbing potential is Mars itself, so resonances of the same order become weaker as we move away from the planet toward the synchronous orbit. Furthermore, resonances with a higher numerator in their ratio (e.g. ``3'' in 3:2 tesseral resonance) depend on the degree 3 spherical harmonics, which are restricted to order 3 and higher of the martian gravity field. Potentials of higher-order harmonics have steeper dependence on distance, and fall off faster with distance from Mars compared to lower-order terms. 

The order of the resonance is also important, as second-order resonances (e.g. 3:1) correspond to Hamiltonian terms that contain the square of the moon's eccentricity and inclination, which makes them weaker if the moon is on a near-circular, near-equatorial orbit. We numerically explored possible effects of the 5:3 at 4.3~$R_M$ and found them to be negligible. Therefore, we will explore only first order tesseral resonances here, and follow them up to degree five, which leaves the 3:2, 4:3 and 5:4 tesseral resonances in consideration.

\begin{figure}[ht]
\epsscale{.6}
\plotone{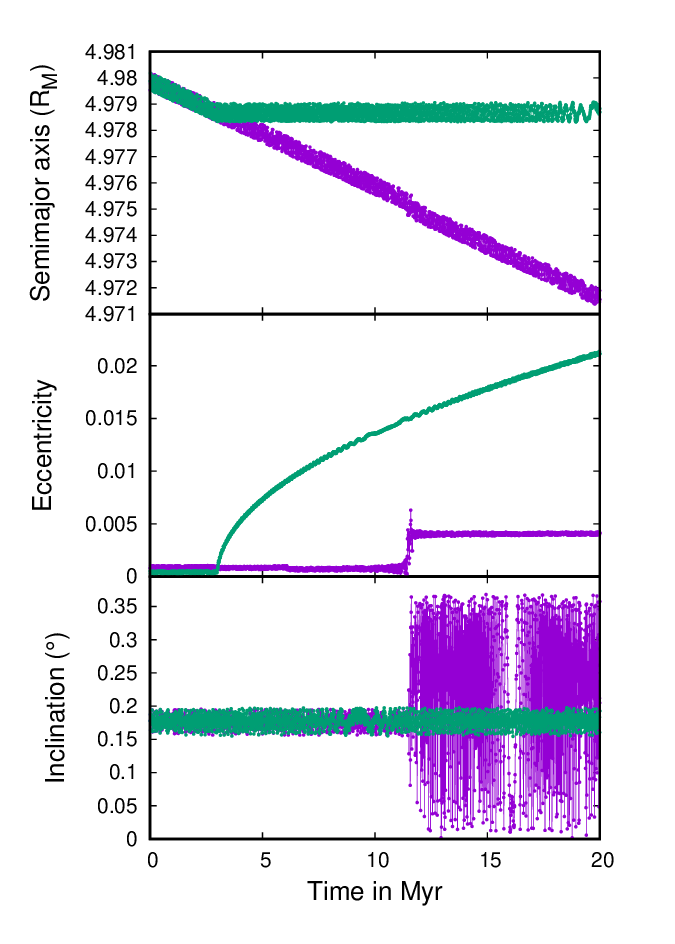}
\caption{Two simulations of Phobos encountering the 4:3 tesseral resonance with martian rotation. The simulation with very low initial eccentricity (green) results in capture into the eccentricity-type tesseral three-body resonances involving Phobos, martian figure and the Sun, while the simulation with initial $e_P=0.001$ (purple) crosses the resonance without capture. The inclination is plotted relative to martian equator and includes a forced inclination of $0.18^{\circ}$. Assuming constant martian $Q$, this resonance happened 1.8~Gyr ago.}
\label{43early}
\end{figure}

Figure \ref{early} shows four different simulations of Phobos's encounter with the 3:2 tesseral resonance at 4.6~$R_M$. We find that even at 100 times the nominal tidal evolution rate there is certain capture of Phobos on an initially circular orbit into a $e$-type T3BR adjacent to the 3:2 tesseral resonance (left hand panel). This T3BR is equivalent to those found close to the 2:1 tesseral resonance, except that this $e$-type T3BR does not break as the eccentricity grows (we confirmed this up to $e_P=0.14$), potentially leading to the sesquinary catastrophe. 

A pre-resonance $e_P=0.002$ is sufficient to avoid capture into this T3BR, and we used this initial $e_P$ in subsequent simulations that use the more realistic migration rate. The right-hand panel in Fig. \ref{early} shows three simulations with identical initial conditions, but with Phobos migrating at 100, 10, and 1 times the nominal tidal evolution rate. All three simulations have final $e_P=0.011$. 

Accelerated simulations have final $i_P=0.2^{\circ}$, but the realistic simulation shows some amount of chaotic behavior during the resonance crossing and results in $i_p=0.55^{\circ}$. We conclude that if Phobos crossed the 3:2 tesseral resonance, it is guaranteed to avoid the capture into $i$-type T3BRs at the 2:1 tesseral resonance. On the other hand, we now require Phobos to have $e_P > 0.002$ going into the 3:2 tesseral resonance to avoid capture into its own three-body harmonics.  

Fig. \ref{43early} shows two simulations of Phobos crossing the 4:3 tesseral resonance at about 4.98~$R_M$. Both simulations used the nominal tidal evolution rate (green line). The initially circular orbit is captured into an $e$-type T3BR much like in Fig. \ref{early}. An initial $e_P=0.001$ is sufficient to avoid capture. When initial $e_P=0.001$ is used, resonance crossing is a simple jump resulting in final $e_P=0.004$ and $i_p=0.2^{\circ}$. The outcome of the 4:3 tesseral resonance guarantees that there is no T3BR capture at 3:2 tesseral resonance, but to avoid very large eccentricities we need $e_P \geq 0.001$ before the 4:3 tesseral resonance.    

\begin{figure}[ht]
\epsscale{.6}
\plotone{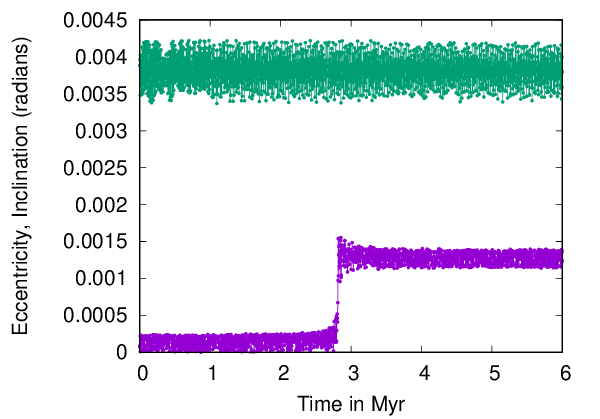}
\caption{Eccentricity (purple) and inclination in radians (green) in simulation of Phobos encountering the 5:4 tesseral resonance with martian rotation. The inclination is plotted relative to martian equator and includes a forced inclination of $0.22^{\circ}$ (0.0038 rad). Assuming constant martian $Q$, this resonance happened 2.5~Gyr ago.}
\label{54early}
\end{figure}

Finally, Fig. \ref{54early} plots a single simulation of 5:4 tesseral resonance crossing at 5.2~$R-M$, conducted at the nominal tidal evolution  rate. Despite very low initial $e_P$ and free $i_p$ (note that the inclination forced by the Sun is $0.22^{\circ}$ at this distance) Phobos avoids capture into any T3BRs, which appear to be very weak for this tesseral resonance. The final $e_P=0.0013$, while the inclination appears unaffected. However, there is a caveat that we use only the $C_5^5$ harmonic to model this resonance, so we likely underestimate the effects on the inclination which are likely to arise from the $C_6^5$ harmonic. Nevertheless, given the pattern of previously explored tesseral resonances, we expect any changes to inclination to be small, below $0.1^{\circ}$, even if these harmonics were included.

We conclude our examination of the more distant tesseral resonances with the conclusion that Phobos could have in principle formed beyond 5.2~$R_M$ on a circular orbit in its Laplace plane, and then gradually accumulated small amounts of eccentricity and inclination by crossing the 5:4, 4:3, and 3:2 tesseral resonances, with each previous resonance introducing enough eccentricity to avoid capture into the $e$-type T3BR associated with the next resonance. Finally, the 3:2 tesseral resonance induces more than enough inclination for Phobos to avoid capture into the $i$-type T3BR associated with the 2:1 tesseral resonance. An important caveat for our model is that we used present-day figure of Mars is our simulations, but geophysical evolution (such as formation of the Tharsis bulge) may have changed martian gravity harmonics over time. We note that this is only relevant for more ancient tesseral resonance crossings discussed in this Section, as 2:1 tesseral resonance would have happened within the last Gyr. Furthermore, even if martian gravity harmonics changed over time, our analysis largely stands as long as these harmonics had approximately similar magnitudes in the past (as we are not interested in their phase). Additionally, this scenario depends on the strength of eccentricity damping within Phobos, as well as on the dependence of martian tidal response on frequency, especially closer to the synchronous orbit. These issues are discussed in the next Section.

\section{Discussion}\label{sec:dis}

We have numerically explored the dynamics of major resonances that have affected Phobos during its tidal evolution. Here we will attempt to formulate complete hypothetical dynamical histories of Phobos that are consistent with our results. While the dynamics of some of the resonances is highly chaotic (this is notably true for the strong 2:1 tesseral resonance), our simulations are adequate to provide us with sufficient understanding of Phobos's orbital history. We first summarize the constraints that we determined in this work in Table~\ref{tab:constraints}
.
\begin{deluxetable}{lccc}
\tablenum{4}
\vspace{-.2in}
\tablecaption{A summary of the constraints found in this work. The orbital elements given are the values required prior to passage through a given resonance.  \label{tab:constraints} }
\tablewidth{0pt}
\tablehead{ 
\colhead{} & \colhead{} & \multicolumn{2}{c}{Pre-resonance constraint}  \\
\colhead{Resonance} & \colhead{$a_P$ $(R_M)$} & \colhead{APH} & \colhead{RPH} 
}
%\decimalcolnumbers
\startdata 
3:1 tesseral & 2.91 & 
    $\begin{array}{rl} 
        e_P & \approx 0.019 \\ 
        i_P & \approx 0.7^\circ  
    \end{array}$ &
    $\begin{array}{rl} 
        e_P & \approx 0.019 \\ 
        i_P & \approx 0.7^\circ 
    \end{array}$ \\
$\begin{array}{l}
    \text{Touma (2:3 solar)}\\
    \hspace{1em} \text{Non-capture} \\
    \rule{0pt}{0.25ex}\\
    \hspace{1em} \text{Capture}\\
\end{array}$ & 
$\begin{array}{c}
    2.94\\
    \rule{0pt}{8ex}\\
\end{array}$
& 
$\begin{array}{c}
    \begin{array}{rl} 
        e_P & \approx 0.018 \\ 
        i_P & \approx 0.8^\circ 
    \end{array}\\
    \text{--}
\end{array}$ 
&
$\begin{array}{c}
    \rule{0pt}{2ex}\\
    \begin{array}{rl} 
        e_P & \approx 0.018 \\ 
        i_P & \approx 0.8^\circ 
    \end{array} \\
    \begin{array}{rl} 
        e_P & \approx 0.025 \\ 
        i_P & \approx 0.25^\circ 
    \end{array}
\end{array}$
\\
Yoder (1:2 solar) & 3.2\phantom{0} & $\begin{array}{rl}&e_P \approx 0.02 \\ & i_P \approx 0.8^\circ \end{array}$ & $\begin{array}{c}\text{Formation interior with outward migration}\\ \text{Mars obliquity}\ \theta_M \ge 45^\circ\end{array}$\\
3:1 with Deimos & 3.33 &No constraints & --\\
2:1 tesseral & 3.8\phantom{0} & $i_P>0.1^\circ$ & --\\
2:1 with Deimos & 4.35 & No constraints & --\\
3:2 tesseral & 4.6\phantom{0} & $e_P>0.002$ & --\\
4:3 tesseral & 4.98 & $e_P>0.001$ & --\\
5:4 tesseral & 5.2\phantom{0} & No constraints & --\\
\enddata
%\tablecomments{}
\end{deluxetable}

\begin{figure}[ht]
\epsscale{1.15}
\plottwo{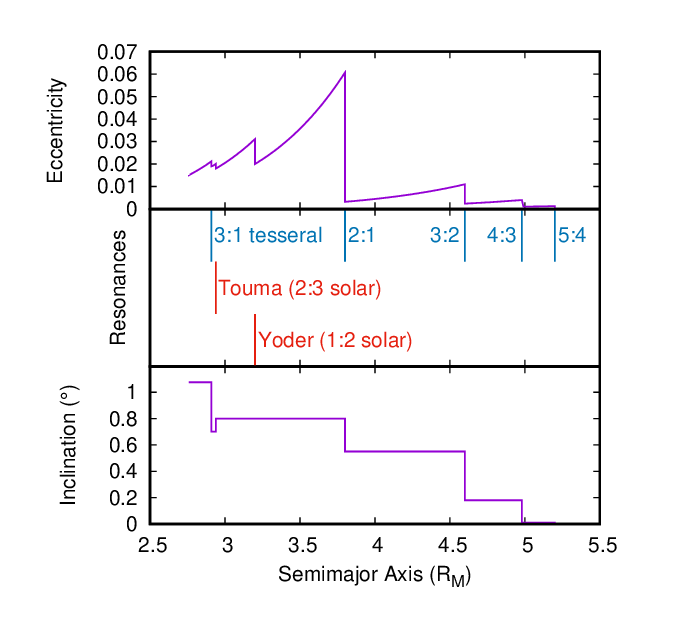}{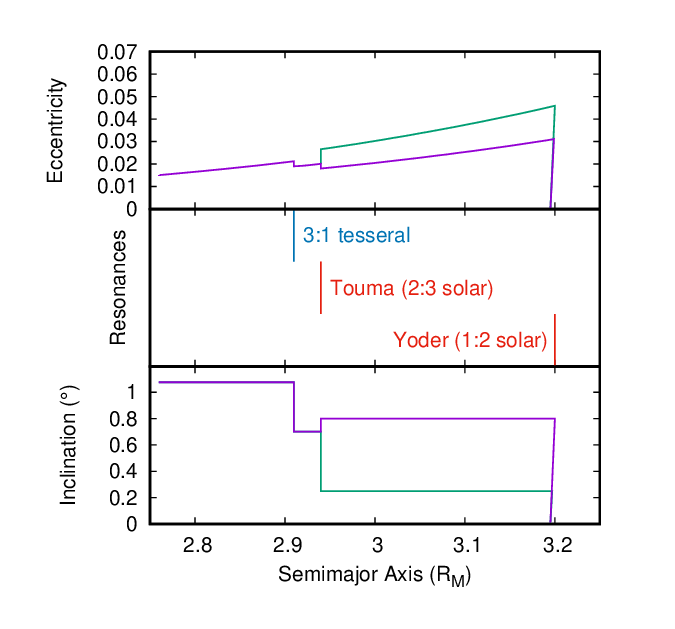}
\caption{The evolution of Phobos's eccentricity and inclination under the Ancient Phobos (left) and Recent Phobos (right) hypotheses. The middle panels indicate the resonance locations. Between resonances we used the relation $e_P \propto a^{6.46}$, based on work of \citet{gol09} on tidal dissipation in rubble piles. In the right-hand panel the teal line plots an alternative evolution path in which Phobos experiences temporary capture (Fig. \ref{32ssimpl}) rather than a ``kick'' (Fig. \ref{32back}) at the Touma resonance at 2.94~$R_M$.}
\label{evo_plots}
\end{figure}

Figure \ref{evo_plots} shows the ``nominal'' reconstructed orbital histories for the APH and RPH. Due to the multiple ways Phobos can cross the Yoder resonance, we are not able to exclude (or confirm) either of these two possibilities. Furthermore, the two ways Phobos can cross the Touma resonance lead to two different orbital histories for young Phobos (purple and teal lines in the right-hand panel of Fig. \ref{evo_plots}). 

The APH curve is most dependent on our assumptions about eccentricity damping within Phobos (and Mars). This is simply the consequence of eccentricity being a power law function of semimajor axis, and ancient Phobos scenarios requiring much more orbital evolution than those proposing a young Phobos. Given that we estimated that an inward-migrating Phobos should enter the Yoder resonance at 3.2~$R_M$ with $e_P \approx 0.02$, and that we assume $e_P \propto a_P^{6.46}$, an ancient Phobos must under these assumptions have $e_P \approx 0.06$ upon exit from the 2:1 tesseral resonance at 3.8~$R_M$. This is close to the largest eccentricity we observed at the end of our simulations of the 2:1 tesseral resonance crossing, and suggests Phobos may have been captured into a tesseral three-body resonance during this event. 

We assume the eccentricity damping has approximately equal contributions from martian constant-$Q$ tides with $e_P \propto a_P^{3.1}$ \citep{yod82} and Phobos's satellite tides with $e_P \propto a_P^{3.36}$ \citep{gol09}. If the tides on Mars follow a constant-lag frequency dependence, the power-law dependence of Phobos's $e$ on its $a$ due to martian tides alone would have an exponent larger than five \citep{yod82}. Combined with realistic dissipation within Phobos, this rapid eccentricity damping would indicate that the current eccentricity of Phobos cannot be explained by the 2:1 tesseral resonance passage, seemingly ruling out the APH. The same would apply to cases in which the dissipation within Phobos is significantly larger (by factor of two at least) than the estimate of \citet{gol09}. \citet{bag21} use seismic data to argue that the quality factor $Q$ of Mars is a very weak function of frequency, so here we used the constant-$Q$ model for simplicity. On the other hand, dissipation within Phobos is not currently constrained by data, and a value that would invalidate APH is certainly within possibility. Therefore, any constraints on the dissipation within Phobos, possibly from the MMX mission, would be most useful for constraining its dynamical history.

The RPH curve covers a much smaller range of semimajor axes and therefore a much shorter temporal history. Not only does the RPH involve smaller and less constraining eccentricity damping, but the proposed origin of eccentricity during the capture in the Yoder resonance can provide significant ($e_P \simeq 0.05$), though poorly-constrained, post-resonance eccentricity. Therefore it would be difficult to realistically rule out the RPH through a measurement of eccentricity damping. 

The requirement that Phobos's initial $e$ is excited while the martian ring is still present does put some constraints on RPH. If Phobos has $a_P=3.2~R_M$ during the Yoder resonance capture, an eccentricity of $e_P=0.05$ would put the pericenter at $3.04~R_M$, requiring a significant gap between the ring and newly-formed Phobos. Some modeling is clearly needed to estimate how eccentric Phobos could become at this stage before colliding with the ring. This is in addition to the wider issue of RPH requiring that any vestige of the ring is removed over the last $10^8$~yr or so, which \citet{mad23} find to be unlikely but, in our opinion, requires direct modeling of the ring dynamics as Phobos migrates from the initial $a_P \simeq 3.2~R_M$ to the current $a_p=2.76~R_P$. 

The teal line in the right-hand panel of Figure \ref{evo_plots} plots an alternative history in which the initial eccentricity of a young Phobos after the Yoder resonance was high, but its inclinations was low, leading to a capture into the Touma resonance (cf. Fig. \ref{32ssimpl}). While this evolution path is less robust than the purple line, which assumes a non-capture crossing of the Touma resonance, this high-$e$, low-$i$ version of Phobos's dynamical history cannot be excluded. The tidal evolution parameters we adopted restrict this alternate evolution path to the RPH, as $e_P=0.05$ after the Yoder resonance may be too high to be the result of subsequent inward passages through the 2:1 tesseral resonance and the Yoder resonance combined, assuming our nominal eccentricity damping. Regardless of the prior evolution and the age of Phobos, the passage through the 3:1 tesseral resonance always produces consistent $e$ and $i$ kicks, as shown in Fig. \ref{31harm}

One additional issue that is affected by martian tidal response as a function of frequency is how close to synchronous orbit an ancient Phobos could have originated. If we use our standard assumptions of constant martian $Q=80$ and $k_2=0.14$, the 5:4 tesseral resonance would have been crossed by Phobos about 2.5 Gyr ago. Different values of $Q$ and $k_2$ for Mars in the literature \citep{jac14} do not change this estimate enough to avoid the crossing of the 5:4 resonance during the lifetime of Mars. However, if the tidal response of Mars were to follow a ``constant lag'' dependence, then Phobos's tidal evolution rate at $5.2~R_M$ would be about eight times slower than in the constant-$Q$ case, and not all of the tesseral resonances discussed in the Section~\ref{sec:early} would be crossed by Phobos over the age of Mars. But, as we note above, seismic data have been used to argue that Martian quality factor $Q$ is a weak function of frequency \citet{bag21}, therefore we argue that the uncertainty about whether Phobos was able to cross more ancient tesseral resonances is not a good argument against the APH.

In the previous section we concluded that the present orbit of Phobos is consistent with its formation a dynamically cold orbit beyond $5.2~R_M$ followed my inward migration. According to our integrations, formation of Phobos with $e=0$ and $i=0$ between the 2:1 and 5:4 tesseral resonances (i.e. $3.8-5.2~R_M$) would result in its capture into one of the long-term stable tesseral three-body resonances accompanying the tesseral resonances. Long-term residence in tesseral three-body resonances would increase Phobos's $e$ or $i$ to the point when a sesquinary catastrophe is likely to occur \citep{cuk23}. 

The most likely outcome of the sesquinary catastrophe is re-accretion on a dynamically cold orbit interior of the resonance, although the numerical work to confirm this is still in progress \citep{ana24}. The resulting low $e$ and $i$ would lead to capture into the a three-body resonance associated with the next inward tesseral resonance, repeating the cycle. Ultimately, Phobos would have re-accreted just interior to 2:1 tesseral resonance at $3.8~R_M$, which is in conflict with constraints from its present $e$ and $i$ (cf. Fig. \ref{yoder_kick}). This argument seemingly excludes Phobos's formation anywhere between $3.2~R_M$ and $5.2~R_M$, but it rests on the assumption there are no significant dynamical effects from objects other than Mars and the Sun. 

It is important to understand that Phobos being ancient does not invalidate the idea that ring-moon cycle happened at Mars, but would just mean that Phobos is not part to that cycle. The ring-moon cycle may have finished billions of years ago \citep[cf.][]{mad23}, and one of the (last?) moons in the cycle could have given Phobos a small eccentricity through a mutual MMR crossing. An induced $e_P=0.001$ acquired between 4:3 and 5:4 resonances would be enough to put Phobos on the right track to potentially match its present orbit. Therefore we need to be cautious when making claims where and when Phobos formed on the basis of presence or absence of very small inclinations and eccentricities. We can only firmly exclude Phobos's (re-)accretion between $3.2~R_M$ and $3.8~R_M$, as Phobos crossed that interval less than 0.5~Gyr ago so it is highly unlikely there were any inner martian moons present other than Phobos, and formation interior to 2:1 tesseral resonance but exterior to Yoder SSR results in $e$ and $i$ that are grossly inconsistent with the present orbit of Phobos (as discussed in Section \ref{sec:yoder}). 

Our main conclusion is that we cannot exclude either the APH or the RPH, and that they are both compatible with the present orbits of Phobos and Deimos, given our current constraints on tidal dissipation in the martian system. In order to rule out one of the two hypotheses, either new observational constraints or a additional analyses are required. As we noted earlier, the viability of APH depends on the eccentricity evolution of Phobos that would follow our nominal estimate of $e_P \propto a_p^{6.46}$. A stronger dissipation within Phobos would not allow enough eccentricity to survive from the dominant 2:1 tesseral resonance to the present. A direct measurement of damping  of either Phobos's eccentricity or its physical librations would help resolve this issue. Additionally, some of the most important past resonances of Phobos like the Yoder resonance (Fig. \ref{yoder_capture}) as well as the long-term dynamics of Deimos (Fig. \ref{lsread_all}) depend on the history of martian obliquity, which is chaotic \citep{war73, tou93, las04}. Precise reconstruction of martian obliquity up to 100~Myr ago may constrain dynamical processes investigated in this paper. Presently, the reconstructed history of martian obliquity reaches only about halfway to this epoch \citep{zee22}. Extending these computations further back in time despite dynamical chaos may require geological data from either Earth or Mars \citep{las11}, with the terrestrial data possibly constraining planetary orbits and the martian record potentially revealing past obliquity values of Mars \citep[cf.][]{hol18}. Finally, samples of Phobos returned by the MMX mission \citep{kur22} may be able to detect or exclude the cosmic ray exposure of Phobos's material that would be expected if it were accreted from a long-lived ring as proposed by RPH \citep{hes17}.

\section{Summary}

In this work we have used direct numerical simulations to revisit the past inward migration of Phobos first explored by \citet{yod82}. While this work relies on certain assumptions about tidal properties of Mars and Phobos, which become more uncertain as we go further back in time, we find that we can make some useful inferences about the past orbit of Phobos. Here is the summary of our main results:

1. On its inward migration Phobos crossed at least two semi-secular resonances with the Sun at 2.94 and 3.2 Mars radii ($R_M$), and a 3:1 resonance with martian rotation at 2.91~$R_M$. Additionally, depending on the formation time of Phobos, it could have crossed a number of other resonances with the rotation of Mars located 3.8 to 5.2~$R_M$.

2. If Phobos formed as a part of a ring-moon cycle about 100 Myr ago \citep{hes17}, its eccentricity and inclination can plausibly be generated from ring-torque-driven outward migration into the 1:2 semi-secular solar resonance at 3.2~$R_M$ (``Yoder resonance'').

3. Alternatively, Phobos could have acquired its eccentricity and inclination from crossings of more ancient resonances with martian rotation as proposed by \citet{yod82}, with the 2:1 tesseral resonance at 3.8~$R_M$ being the most important. 

4. We find that resonances with martian rotation are surrounded by previously undiscovered ``tesseral three-body resonances'' involving martian rotation, the Sun and Phobos. While tidal capture into resonances with martian rotation is not possible, capture into tesseral three-body resonances is observed in our simulations, putting additional constraints on the dynamical history of Phobos.

5. We find that the orbit of Deimos is somewhat chaotic due to planetary perturbations, especially during epochs of high martian obliquity. The resulting stochastic variations in Deimos's eccentricity remove the possibility of constraining the past Phobos-Deimos resonances on the basis of the current orbit of Deimos.

6. We find that with the available data we cannot use the Phobos's present orbit to determine its age. Phobos could have formed less than 100~Myr ago from the outer edge of circum-martian disk at 3.2~$R_M$ \citep{hes17}, or could have formed much further out, but interior to synchronous orbit ($5.2~R_M < a < 6~R_M$) billions of years ago.   

We hope that future potential measurements of dissipation within Phobos, as well as improved understanding of the history of martian obliquity could help us potentially rule out one of these hypotheses. In future work, we plan to explore whether a circum-martian ring could have been dynamically removed within the last 100~Myr, as required by the ring-moon hypothesis. 

\begin{acknowledgments}
We would like to thank two anonymous reviewers whose comments greatly improved the manuscript. This work was funded by NASA Emerging Worlds Program award 80NSSC23K1266. M\'C would like to dedicate this paper to the memory of Joe Burns.
\end{acknowledgments}

\bibliography{martian_refs}{}
\bibliographystyle{aasjournal}



\end{document}